\documentclass[useAMS,usenatbib]{mn2e}
\usepackage{natbib}
\usepackage{psfig}

\def\gs{\mathrel{\raise0.35ex\hbox{$\scriptstyle >$}\kern-0.6em
\lower0.40ex\hbox{{$\scriptstyle \sim$}}}}
\def\ls{\mathrel{\raise0.35ex\hbox{$\scriptstyle <$}\kern-0.6em
\lower0.40ex\hbox{{$\scriptstyle \sim$}}}}
\def\ls{\mathrel{\hbox{\rlap{\hbox{\lower4pt\hbox{$\sim$}}}\hbox{$<$}}}}
\def\gs{\mathrel{\hbox{\rlap{\hbox{\lower4pt\hbox{$\sim$}}}\hbox{$>$}}}}

\def\msun{{\rm\,M_\odot}}

\title[The Effect of Local Galaxy Surface Density on Star Formation for HI selected galaxies]
      {The Effect of Local Galaxy Surface Density on Star Formation for HI selected galaxies}
\author[M.~T.~Doyle et al.]
       {M.~T.~Doyle,$^{\star}$ and M.~J.~Drinkwater 
        \vspace*{1mm}\\
	$^{\star}$ mtdoyle@physics.uq.edu.au \\
  Department of Physics, University of Queensland, Brisbane, QLD 4072 Australia\\}

\date{\fbox{\sc Draft: \today\ --- Do Not Distribute}}
\date{Accepted 2006 August 2; Received 2006 July 7; in original 2006 March 13}

\pagerange{000--000}

\begin{document}

\maketitle

\begin{abstract}

We present the result of investigations into two theories to explain the star formation rate-density relationship. For regions of high galaxy density, either there are fewer star forming galaxies, or galaxies capable of forming stars are present but some physical process is suppressing their star formation. We use  HI Parkes All Sky Survey's (HIPASS) HI detected galaxies and infrared and radio fluxes to investigate star formation rates and efficiencies with respect to local surface density.  For nearby (vel$<$10000 km s$^{-1}$) HI galaxies we find a strong correlation between HI mass and star formation rate. The number of HI galaxies decreases with increasing local surface density. For HI galaxies (1000$<$vel$<$6000 km s$^{-1}$) there is no significant change in the star formation rate or the efficiency of star formation with respect to local surface density. We conclude the SFR-density relation is due to a decrease in the number of HI star forming galaxies in regions of high galaxy density and not to the suppression of star formation.

\end{abstract}


\section{Introduction} 
  \label{Sec:Paper2Introduction}
\citet{dressler1980b}, finds a well--defined relationship between local galaxy density and morphology for galaxies in clusters, the `morphology--density relation'. Several later studies, e.g. \citet{hashimoto1998}, \citet{goto2003}, \citet{sodre2003} and \citet{nuijten2005} use large optical surveys and 2 or 3D local galaxy density to investigate the Morphology-density relation. They confirm \citet{dressler1980b}'s cluster results: as local galaxy density increases, there is a decrease in the relative number of spiral galaxies and an increase in the number of elliptical and SO galaxies.  

Two studies investigating the morphology-density relation with respect to the star formation rate (SFR) and local galaxy density (hereafter the SFR-density relation) are \citet{lewis2002} and \citet{gomez2003}. These studies are based on optical observations with galaxy SFRs calculated from H$\alpha$ fluxes. To determine the local galaxy density \citet{gomez2003} and \citet{lewis2002} use the local surface density. \citet{lewis2002} uses the SFR for a given density range for clusters compared to field galaxies.  \citet{gomez2003} uses two samples, one sample is separated into field and group/clusters galaxies and the other is the Sloan Digital Sky Survey.  Both \citet{lewis2002} and \citet{gomez2003} conclude the SFR for a given density range decreases as the local surface density increases. \citet{balogh2004} investigate the relationship between the distribution of galaxies and local density and find that star-forming and quiescent (elliptical) galaxies vary with local density.  However, they find the `distribution of H$\alpha$ equivalent widths' of the star-forming population does not depend strongly on environment.

Our work is based on the HI Parkes All Sky Survey Optical Catalogue (HOPCAT), \citep{doylemt2005mnras}. HOPCAT is, at the present time, the largest optically matched HI radio detected catalogue of the whole southern sky. For the first time we investigate the SFR-density relation using a large sample of galaxies containing the basic fuel for star formation, HI. For regions of high galaxy density we ask, `is the decrease in SFR due to fewer star forming galaxies or are galaxies capable of forming stars present but some physical process is suppressing their star formation?'  

When investigating the SFR-density relation, various 3D methods to determine the local galaxy density are used \citep{allam1999, martinez2002, sodre2003} which require redshifts. At the time of writing, no redshift catalogue of the whole of the southern sky exists, hence we use the local surface density, $\Sigma_{10}$, similar to \cite{dressler1980b}, \cite{lewis2002} and \cite{gomez2003}.

To calculate the SFR, \citet{hopkins2001}, \citet{cardiel2003}, \citet{lewis2002} and \citet{gomez2003} use OII, OIII, H$\alpha$, H$\beta$, H$\delta$ and UV detected fluxes, however these bands require correction for dust extinction.  Infrared and radio detections, both indicators of current star formation, are not affected by dust \citep{hopkins2003, cardiel2003} and studies show a correlation exists between IR and 1.4-GHz SFRs \citep{hopkins2002, cardiel2003, yun2001}. In our study we use 843-MHz and 1.4-GHz radio, and the 12, 25, 60 and 100 $\mu$m infrared (IR) detected fluxes to calculate the SFR per HI galaxy. 

The star formation efficiency (SFE) is used to determine the efficiency with which the HI present is converted into stars per year. From the HI galaxy SFR, SFE and $\Sigma_{10}$ we  explain the reason behind the SFR-density relation.    

In Section \ref{Sec:Paper2LocalSurfaceDensity} we describe the process taken to determine the local surface density for each HOPCAT target galaxy.  Section \ref{Sec:Paper2SFRSFE} details the resources we use to obtain the fluxes and the equations to calculate the SFRs and SFEs. We present our results in Section \ref{Sec:Paper2Results} and discuss these results in Section \ref{Sec:Paper2Discussion}.  Section \ref{Sec:Paper2Summary} summarises our work. We use H$_{\circ}$=71 km s$^{-1}$ Mpc$^{-1}$ throughout.

\section{Local Surface Density} \label{Sec:Paper2LocalSurfaceDensity}

To investigate the SFR-density relation we require a measure of local galaxy density. Using the positions for optically matched HOPCAT galaxies and two optical background catalogues, the local surface density for each galaxy is calculated.   In the following sections we introduce the target galaxies from the largest optically matched HI survey of the whole southern sky, HOPCAT.  We also discuss the background catalogues, the SuperCOSMOS Galaxy Catalogue\footnote{SuperCOSMOS Galaxy Catalogue is supplied directly from Mike Read at Royal Observatory Edinburgh} \citep{hambly2001a, hambly2001b, hambly2001c} and the 6dFGS input catalogue \citep{jonesHeath2004} that are used, along with HOPCAT, to calculate the local surface density, $\Sigma_{10}$.

\subsection{HOPCAT Target Galaxies \label{SubSec:Paper2HopcatForDensity}}

Our base sample is optically matched HOPCAT galaxies from match categories 60, 50, and 40, \cite{doylemt2005mnras}.  These HI-optical galaxy match categories are; 60s - matched using independent velocity confirmation, 40s - matched using independent velocity confirmation but multiple galaxies with similar velocities are within the 15$\times$15 arcmin image and 50s - matched without independent velocity confirmation but the chosen galaxy is the only galaxy visible within the image (for a more detailed description of HOPCAT and the match categories please refer to \cite{doylemt2005mnras}). As these are the most reliable match categories we use their velocities in the local surface density calculations.   In the subset of HOPCAT analysed in this paper the match category breakdown is 72 per cent 60s, 24 per cent 40s and 4 per cent 50s.

The basis of HOPCAT is the HI All Sky Survey catalogue (HICAT), \citep{meyer2004, zwaan2004}.  This is a flux limited catalogue due to the sensitivity of the Parkes 21-cm multibeam receiver. Due to this, and the magnitude limits of the two background galaxy catalogues (see Section \ref{SubSec:Paper2BackgrCats}), we only use galaxies between 1000 and 6000 km s$^{-1}$ for the SFR, SFE and $\Sigma_{10}$ analysis.  
	
\subsection{Background catalogues for Density Calculations \label{SubSec:Paper2BackgrCats}} 

We use two whole southern sky catalogues, SuperCOSMOS Galaxy Catalogue and The 6dFGS' \ 2MASS input catalogue to calculate the local surface density. 

\begin{figure}
\centerline{\psfig{file=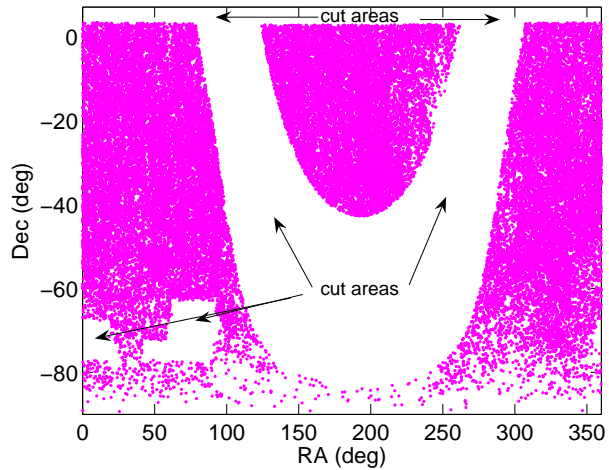,angle=0,width=3.5in}}
  \caption{\small{SuperCOSMOS Galaxy Catalogue sky coverage.  Note the galactic plane ($|b|<20^{\circ}$), the Large and Small Magellanic Clouds and Dec.$>$+3.25 are cut areas. Any HOPCAT target galaxy with less than 50 per cent of the 10th neighbour area overlaps a cut area, have corrected densities. Any HOPCAT target galaxy within a cut area or with greater than 50 per cent of its 10th neighbour area within a cut area, are removed from our target lists.}}
  \label{Fig:Paper2SCosAllSkyPlot}
\end{figure}

The SuperCOSMOS Galaxy Catalogue (hereafter SCosGC), a subset of the SuperCOSMOS Sky Survey which contains only visually confirmed galaxies, is supplied directly by Mike Read from the Royal Observatory Edinburgh. SCosGC contains approximately 107000 objects, has a faint magnitude limit of B$_j$ $<$ 17.0 mag and covers the whole of the southern sky up to +3.2$^{\circ}$ Dec. The areas around the galactic plane ($|b|<20^{\circ}$) and the Large and Small Magellanic Clouds are not included, Fig. \ref{Fig:Paper2SCosAllSkyPlot}.  Read (Private Communication) compares SCosGC with the APM catalogue and finds a completeness $>$ 95 per cent.  The resulting densities calculated using SCosGC are referred to as HOPCAT-SCosGC.

\begin{figure}
\centerline{\psfig{file=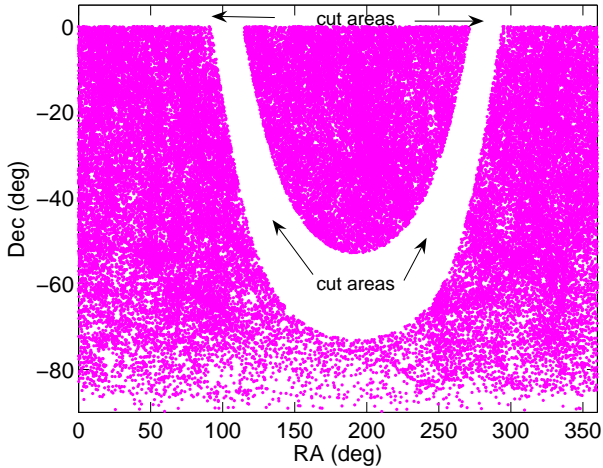,angle=0,width=3.5in}}
  \caption{\small{6dF2MASS Galaxy Catalogue sky coverage.  Note the the galactic plane ($|b|<10^{\circ}$) and Dec.$>$0 are cut areas. Any HOPCAT target galaxy with less than 50 per cent of the 10th neighbour area overlaps a cut area, have corrected densities. Any HOPCAT target galaxy within a cut area or with greater than 50 per cent of its 10th neighbour area within a cut area, are removed from our target lists.}}
  \label{Fig:Paper2K6dFAllSkyPlot}
\end{figure}

The 6dFGS' 2MASS input catalogue (hereafter 6dF2MASS) is obtained from the 6dFGS web site (http://www.mso.anu.edu.au/6dFGS/).  These objects are detected in the near-infrared with a K-band faint magnitude limit of {\it{K}} $<$ 12.75 mag. 6dF2MASS covers the whole of the southern sky up to -00$^{\circ}$ Dec. and contains approximately 121000 objects with the galactic plane ($|b|<10^{\circ}$) not included, Fig. \ref{Fig:Paper2K6dFAllSkyPlot}. A conversion of $<${\it{$B_j$}}-{\it{K}}$>$=4 mag, calculated from galaxies common to SCosGS and 6dF2MASS, is used to convert 6dF2MASS' {\it{K}} to {\it{B$_j$}} mags to standardise all magnitudes in this work.  The densities calculated using 6dF2MASS are referred to as HOPCAT-6dF2MASS.

\subsection{Density Calculation \label{SubSec:Paper2DensityCalc}}

We use local surface density, $\Sigma_{10}$, similar to \citet{dressler1980b}, \citet{lewis2002} and \citet{gomez2003}, to calculate the local galaxy density. The area of density is calculated from a Mpc radius (r) between the position of the HOPCAT target galaxy and the 10th nearest neighbouring galaxy using SCosGC and 6dF2MASS galaxy positions.  As this area (hereafter 10th neighbour area) includes a total of 11 galaxies, the $\Sigma_{10}$ density is calculated as follows:

\begin{equation}
\Sigma_{10}(\textrm{Mpc}^{-2})=\frac{11}{r\pi^2}. \label{eq:Paper2Sigma10}
\end{equation}

As $\Sigma_{10}$ is based on a 2D view of the sky, fore and background galaxies can falsely inflate the density value.  \citet{dressler1980b}, \citet{lewis2002} and \citet{gomez2003} use a faint absolute magnitude limit equivalent to M$_{B_j}$=-17 mag to eliminate the effect from background galaxies.  As we are using nearby HI detected galaxies, where fainter galaxies are detected, we require a fainter absolute magnitude limit as well as a foreground limit. We use M$_{B_j}$=-21 and M=-17 mag for arbitrary bright and faint absolute magnitude limits respectively. The corresponding bright and faint apparent magnitude limits are calculated using the HOPCAT target galaxy's velocity and only galaxies within this range in SCosGC and 6dF2MASS are used to determine the 10th nearest neighbouring galaxy.

Edge-affected regions occur near the areas cut from SCosGC and 6dF2MASS, Figs. \ref{Fig:Paper2SCosAllSkyPlot} and \ref{Fig:Paper2K6dFAllSkyPlot}.  To reduce this problem we use a number of culling steps. Any HOPCAT target galaxies in edge-affected regions, where less than 50 per cent of the 10th neighbour area overlaps a cut area, have corrected densities. Any HOPCAT target galaxy within a cut area or with greater than 50 per cent of its 10th neighbour area within a cut area, is removed from our target lists. Also any galaxies, mainly those with small velocities, having fewer than 10 neighbouring galaxies are removed from the target lists. HOPCAT-SCosGC and HOCAT-6dF2MASS target lists contain 2126 and 2603 galaxies respectively for 1000$<$vel$<$6000 km s$^{-1}$.

\section{Star Formation Rates \& Efficiency} \label{Sec:Paper2SFRSFE}

To investigate star formation and hence the SFR-density relation, the HI galaxies SFRs and SFEs are calculated. The SFR is the rate of star formation in solar masses per year. SFE is the efficiency at which the HI present in galaxies is converted into stars per year and is independent of distance. We use the two HOPCAT target lists, HOPCAT-SCosGC and HOPCAT-6dF2MASS, to search for IR and radio fluxes to calculate the SFR and SFEs. Hereafter the 4 data sets will be referred to as shown in Table \ref{Tab:Paper2DataSetNames}.

In the following section we discuss the IR and radio data resources to obtain fluxes and the equations to calculate the SFR and SFE.  We also discuss the difference between the IR and radio SFRs. 

\subsection{SUMSS and SN6I Catalogues \label{SubSec:Paper2SumssSN6ICats}}
To calculate SFR and SFE we utilise 2 sources of radio data, The Sydney University Molonglo Sky Survey (SUMSS) \citet{mauch2003} and a catalogue of SUMSS, NRAO VLA Sky Survey (NVSS), 6dFGS and {\it{IRAS}} data (hereafter SN6I) compiled by Tom Mauch \citep{mauch2005PhD}. SUMSS is detected at 843-MHz and has a sky coverage of -90$<$Dec.$<$-30. NVSS is detected at 1.4-GHz and has a sky coverage of -40$<$Dec.$<$0.

To determine the correct SUMSS and SN6I object corresponding to the HOPCAT target galaxies, we use the following analysis to find the optimum search radius for matching the respective catalogues. We compare the number of SUMSS and SN6I galaxies matched to HOPCAT using the original galaxy positions with those matching a `random' catalogue made with the original positions offset by 0.25 degs.  The optimum search radius we use is where the number of (original position) matches  converges to the `random' level defined by the offset positions. The optimum search radius is 30 arcsec with a reliability of a match and completeness (the percentage of possible genuine matches missed) of 80 and 1.24 per cent respectively. Final target galaxy numbers with SUMSS and SN6I fluxes are listed in Table \ref{Tab:Paper2FinalGalNumber}.

The equations we use to calculate the SFR \citep{cram1998} and SFE, from the SUMSS and SN6I fluxes, are:

\begin{equation}
SFR_{Radio}(\msun \textrm{yr}^{-1})=\frac{L_{1.4-GHz}}{2.3\times10^{22}\textrm{WHz}^{-1}}, \label{eq:Paper2SumssTomSFR}
\end{equation}

\begin{equation}
SFE_{Radio}(\textrm{yr}^{-1})=\frac{SFR_{1.4-GHz}}{M_{HI}}.\label{eq:Paper2SumssTomSFE} 
\end{equation}

The uncertainty for the 1.4-GHz radio SFR equation is a factor of 10$^{0.2}$, as discussed by \citet{cram1998}, and is associated with the luminosity calculation.  There are also random uncertainties in the SUMSS flux as listed in the SUMSS catalogue, and an uncertainty of 0.1 per cent in SN6I fluxes as described in \citet{mauch2005PhD}. The total $\Delta$SFR and $\Delta$SFE are:

\begin{equation}
\Delta SFR=SFR \sqrt{\frac{\Delta F}{F}^2+\frac{2\Delta Vel}{Vel}^2+\frac{\Delta SFR _{eq}}{SFR}^2},\label{eq:Paper2DeltaSFR} 
\end{equation}

\begin{equation}
\Delta SFE=SFE \sqrt{\frac{\Delta F}{F}^2+\frac{\Delta S_{int}}{S_{int}}^2+\frac{\Delta SFR}{SFR}^2}, \label{eq:Paper2DeltaSFE} 
\end{equation}

\noindent where F is the flux density, $S_{int}$ is the HI integrated flux and $\Delta SFR_{eq}$ is the uncertainty in the 1.4-GHz SFR equation. As the SFR equation is for 1.4-GHz detected fluxes, the SUMSS fluxes are adjusted using a mean radio spectral index calculated as described in \citet{mauch2005PhD}, consistent with previous spectral index calculations for frequencies below 1.4-GHz \citep{mauch2003, oort1988, hunstead1991, deBreuck2000}. The radio spectral index is defined as a power-law:

\begin{equation}
F_{1.4-GHz}=F_{843-MHz}(1400/843)^\alpha,
 \label{eq:Paper2SpectralIndexTom}
\end{equation}

\noindent where F is the flux density, $\alpha$=-0.7 is the radio spectral index and (1400/843) is the frequency ratio.

\subsection{{\it{IRAS}} \label{SubSec:Paper2IRAS} }

We obtain our second source of fluxes from the Infrared Astronomical Satellite ({\it{IRAS}})\footnote{This research has made use of the NASA/ IPAC Infrared Science Archive, which is operated by the Jet Propulsion Laboratory, California Institute of Technology, under contract with the National Aeronautics and Space Administration.}. We use the {\it{IRAS}} Galaxies and Quasars Catalogue (IRASGQCat) which covers the whole sky. The same method we use to determine the search radius for the SUMSS and SN6I objects is applied. The optimum search radius of 60 arcsec between the HOPCAT and {\it{IRAS}} objects with a reliability and completeness of a match of 97 and 2.7 per cent respectively. Using this radius we search the {\it{IRAS}} web based batch search facility, GATOR (http://irsa.ipac.caltech.edu/cgi-bin/Gator/nph-dd), using our HOPCAT target galaxy positions to obtain flux measurements. Final target galaxy numbers with {\it{IRAS}} fluxes are listed in Table \ref{Tab:Paper2FinalGalNumber}.

The {\it{IRAS}} 12, 25, 60 and 100 $\mu$m fluxes are combined to calculate a total luminosity for each galaxy from which the SFR \citep{cardiel2003} and SFE are calculated as follows:


\begin{eqnarray}
F_{Tot.IR}(\textrm{ergs}~\textrm{cm}^{2}~s)=1.8\times10^{11}&&  \nonumber \\
(13.48F_{12\mu}+5.16F_{25\mu}+2.58F_{60\mu}+F_{100\mu}),\nonumber \label{eq:Paper2IrasFlux}\\
\end{eqnarray}
 
\begin{equation}
SFR_{Tot.IR}(\msun \textrm{yr}^{-1})=4.5\times10^{-44} L_{Tot.IR}\:(\textrm{ergs}\:\textrm{s}^{-1}),
\label{eq:Paper2IrasSFR}
\end{equation}

\begin{equation}
SFE_{Tot.IR}(\textrm{yr}^{-1})=\frac{SFR_{Tot.IR}}{M_{HI}}. \label{eq:Paper2IrasSFE} 
\end{equation}

We use only the uncertainty, $\pm$ 30 per cent,  associated with the SFR equation from \citet{cardiel2003} as discussed in \citet{kennicutt1998}, as any random uncertainties are insignificant in comparison.

%
\begin{table*}
\begin{center}
\medskip
\caption{\small{Data Set Names.
}}
\begin{tabular}{lll}
\hline
Fluxes& HOPCAT-SCosGC & HOPCAT-6dF2MASS  \\
\hline 
{\it{IRAS}}   & HOPCAT-SCosGS-IR  		& HOPCAT-6dF2MASS-IR \\
SUMSS \& SN6I  & HOPCAT-SCosGS-1.4GHz  & HOPCAT-6dF2MASS-1.4GHz  \\
\hline
\end{tabular}
\label{Tab:Paper2DataSetNames}
\end{center}
\end{table*}

\begin{table}
\begin{center}
\medskip
\caption{\small{Galaxy samples. 
}}
\begin{tabular}{lcc}
\hline
Fluxes& HOPCAT-SCosGC & HOPCAT-6dF2MASS  \\
\hline 
SFR vs M$_{HI}$ & ($vel<$10000 km s$^{-1}$)&\\
{\it{IRAS}}      & 1405 &  1405 \\
SUMSS     &340  & 340  \\
SN6I      & 221 & 221   \\
\hline
$\Sigma_{10}$ Plots &( 1000$<$vel$<$6000 km s$^{-1}$)&\\
{\it{IRAS}}      & 923  &  1180 \\
SUMSS     & 172  & 275  \\
SN6I      & 167  & 203   \\
\hline
\end{tabular}
\label{Tab:Paper2FinalGalNumber}
\end{center}
\end{table}

\subsection{Comparing IR and 1.4-GHz Radio based SFRs \label{SubSec:Paper2CompareIRandRadio} }

The SFRs from IR and 1.4-GHz radio are based on two different flux producing mechanisms. The IR flux is a measure of the dust and gas heating during the star formation process and 1.4-GHz radio flux is a measure of sychrontron emission.  Both are measures of current star formation and although they are from different mechanisms, the dust transparent nature of both measurements and the linear correlation between 1.4-GHz radio and IR-based SFR, ensure both are equally reliable measures of current star formation.  However as the uncertainties for IR are $\pm$ 30 per cent, the 1.4-GHz SFRs are more reliable.

\section{Results} \label{Sec:Paper2Results}
In this section we present the results for our HI mass (M$_{HI}$) and SFRs investigations for HOPCAT galaxies. We also discuss the difference between HI radio and optically detected galaxy samples and how this relates to SFRs.  The SFRs and SFEs with respect to $\Sigma_{10}$ results are also presented. Please note: all scatter plots use least-squares line of best fit, calculated using the logs of M$_{HI}$, SFR or SFE, with $\pm$3$\sigma$ slope uncertainty lines. 

\subsection{HI mass-SFR Relationship \label{SubSec:Paper2ResultMassHISFR}}

As HOPCAT is the largest optically matched HI radio detected catalogue, we comprehensively investigate the relationship between M$_{HI}$ and SFR. 

We use the integrated flux (S$_{int}$) and the velocity (vel) to calculate the M$_{HI}$ as follows:

\begin{equation}
M_{HI}(\msun)=2.356\times10^5 \left(\frac{vel}{H_o}\right)^2 S_{int}. \label{eq:Paper2MassHI} 
\end{equation}

The $\Delta M_{HI}$, resulting from the velocity and S$_{int}$ \citep{zwaan2004}, is:

\begin{equation}
\Delta vel=1.0\times10^4S_p^{-2}(\textrm{mJy})+5,
\label{eq:Paper2DeltaVel}
\end{equation}

\begin{equation}
\Delta S_{int}=0.5S_{int}^{1/2},
\label{eq:Paper2DeltaSint}
\end{equation}

\begin{equation}
\Delta M_{HI}=M_{HI} \sqrt{\frac{2\Delta vel}{vel}^2+\frac{\Delta S_{int}}{S_{int}}^2}. \label{eq:Paper2DeltaMassHI} 
\end{equation}

The $\Delta$vel depends only on the peak flux S$_p$ \citep{zwaan2004}. 

\begin{figure*}
\centerline{\psfig{file=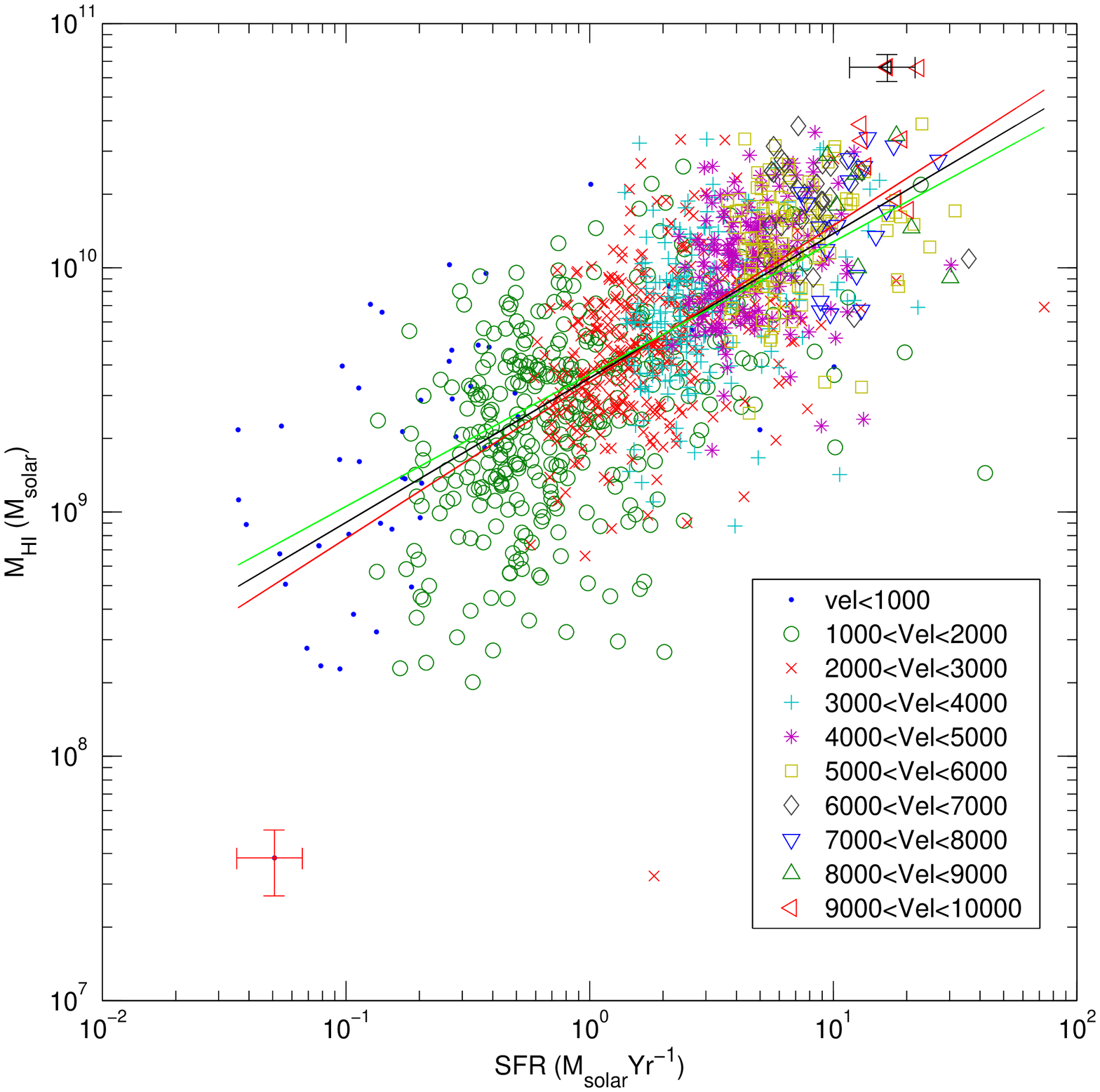,angle=0,width=3.5in}\psfig{file=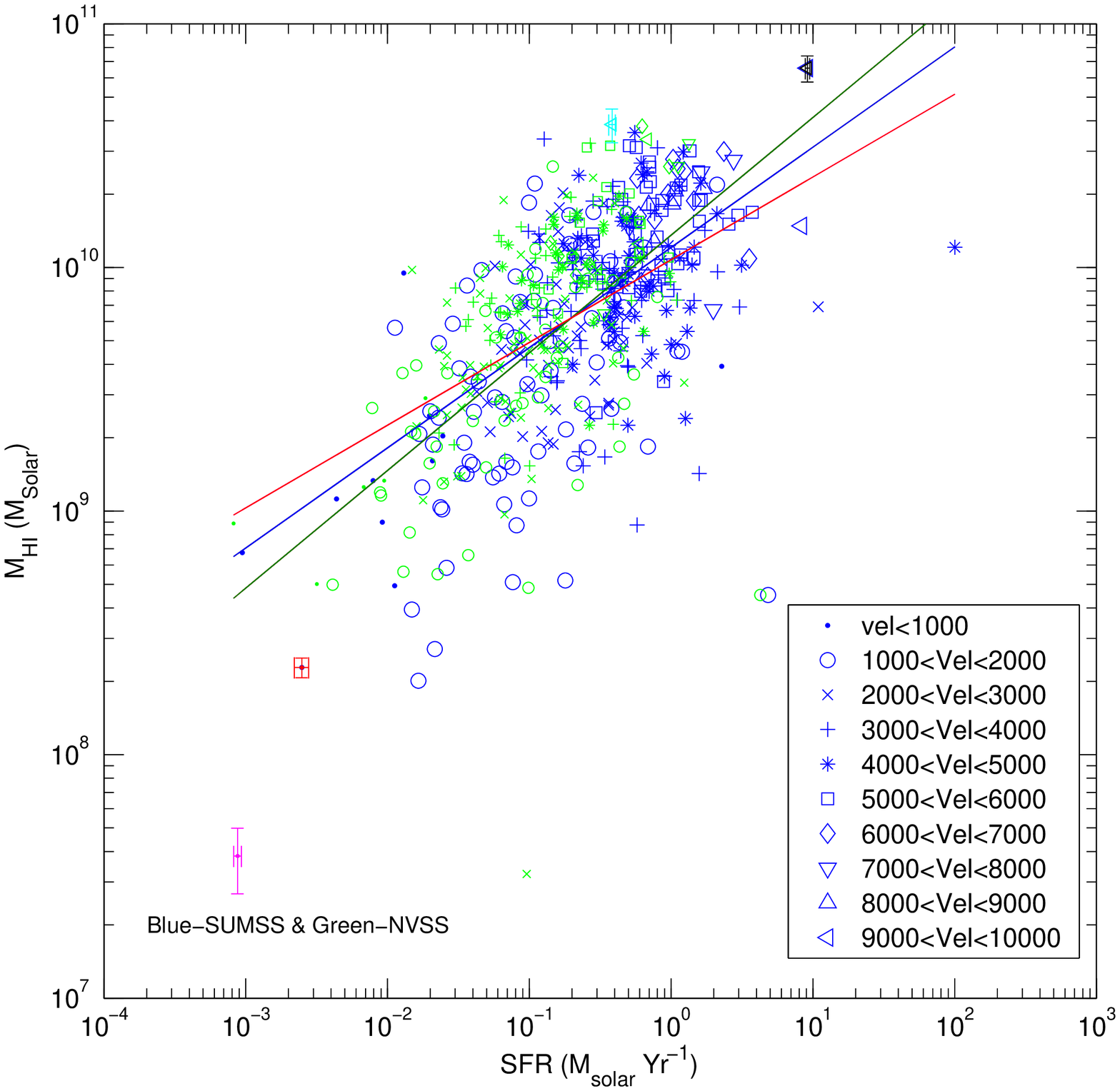,angle=0,width=3.5in}}
  \caption{\small{M$_{HI}$ versus SFR. Least-squares line of best fit, calculated using the logs of M$_{HI}$ and SFR. The gradient uncertainty lines are 3$\sigma$. Left panel: {\it{IRAS}} infrared data with the minimum (red) and maximum (black) uncertainties. Right panel: SUMSS \& NVSS 1.4-GHz radio data with the SUMSS minimum (red) and maximum (black) and NVSS minimum (magenta) and maximum (cyan) uncertainties are shown.   M$_{HI}$ increases with increasing SFR.}}
  \label{Fig:Paper2MassHIVsSFR}
\end{figure*}

We calculate M$_{HI}$ for all HOPCAT target galaxies with SFRs out to 10000 km s$^{-1}$ (IR: 1405 galaxies and 1.4-GHz: 561 galaxies). Fig. \ref{Fig:Paper2MassHIVsSFR} shows the relationship between M$_{HI}$ and SFR with least-squares lines of best fit, as described earlier. $\Delta M_{HI}$ are shown in Fig. \ref{Fig:Paper2MassHIVsSFR} as; IR, red (min) and black (max); 1.4-GHz, Sumss, red (min) and black (max) and NVSS, magenta (min) and cyan (max). It is seen for both IR and 1.4-GHz, the M$_{HI}$ increases with increasing SFR.  

However the flux-limited nature of our data is seen in Fig. \ref{Fig:Paper2MassHIVsSFR} in the colour/symbol velocity groupings. For a given velocity only galaxies down to the sensitivity of the receiver are detected.  To reduce this effect we analyse M$_{HI}$ versus SFR for 1000 km s$^{-1}$ velocity ranges and fit each with a line of best fit as described above.  These results also show an increase in M$_{HI}$ with increasing SFR. 

\begin{figure}
\centerline{\psfig{file=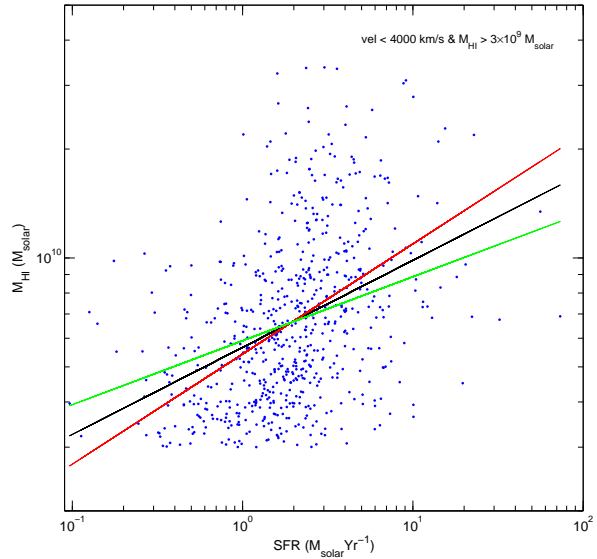,angle=0,width=3.5in}}
  \caption{\small{M$_{HI}$ versus Star formation rate for volume limited sample vel $<$ 4000 km s$^{-1}$ and M$_{HI}$ $>$ $3\times10^{9} M_{\odot}$ for {\it{IRAS}} infrared data. Least-squares line of best fit, as described in Fig. \ref{Fig:Paper2MassHIVsSFR}, are fitted. M$_{HI}$ increases with increasing SFR.}}
  \label{Fig:Paper2MassHIVsSFRVolumeLimited}
\end{figure}

Though Fig. \ref{Fig:Paper2MassHIVsSFR} and the velocity separated analysis show M$_{HI}$ increases with increasing SFR, the flux-limited nature of our sample may still affect results.  To reduce the effect we use a volume limited subsample of our HOPCAT target list. We use only galaxies out to 4000 km s$^{-1}$ with M$_{HI}$ greater than $3\times10^{9} M_{\odot}$ (see fig. 2 of \citet{zwaan2005} for velocity and M$_{HI}$ limits). Fig. \ref{Fig:Paper2MassHIVsSFRVolumeLimited} shows the M$_{HI}$-SFR relation for the volume limited sample. Again we use least-squares line of best fit. The M$_{HI}$ still increases with increasing SFR. The gradients (m) in Fig. \ref{Fig:Paper2MassHIVsSFR} are 0.6$\pm$0.02 and 0.4$\pm$0.02 for IR and 1.4-GHz respectively and m=0.3$\pm$0.02 for the volume limited sample, Fig. \ref{Fig:Paper2MassHIVsSFRVolumeLimited}. 

Though the gradient for the volume limited sample is less steep, again there is a clear increase in M$_{HI}$ with increasing SFR. We conclude, there is a strong correlation between a galaxy's M$_{HI}$ content and SFR.  

%
%

\subsection{Comparing HI radio and Optical galaxy samples with respect to local surface density \label{SubSec:Paper2ResultsCompareHiOp}}

We use the 6dFGSs Second Data Release (hereafter 6dFGS2DR), obtained using their web access page\footnote{http://www-wfau.roe.ac.uk/6dfgs/form.html} and our HOPCAT target galaxies to compare HI and optically detected galaxy samples for the southern sky.  We have limited the 6dFGSDR2 data to contain only galaxies with the same velocity limit as our HOPCAT target list (1000$<$vel$<$6000 km s$^{-1}$). 

\begin{figure*}[b!]
\centerline{\psfig{file=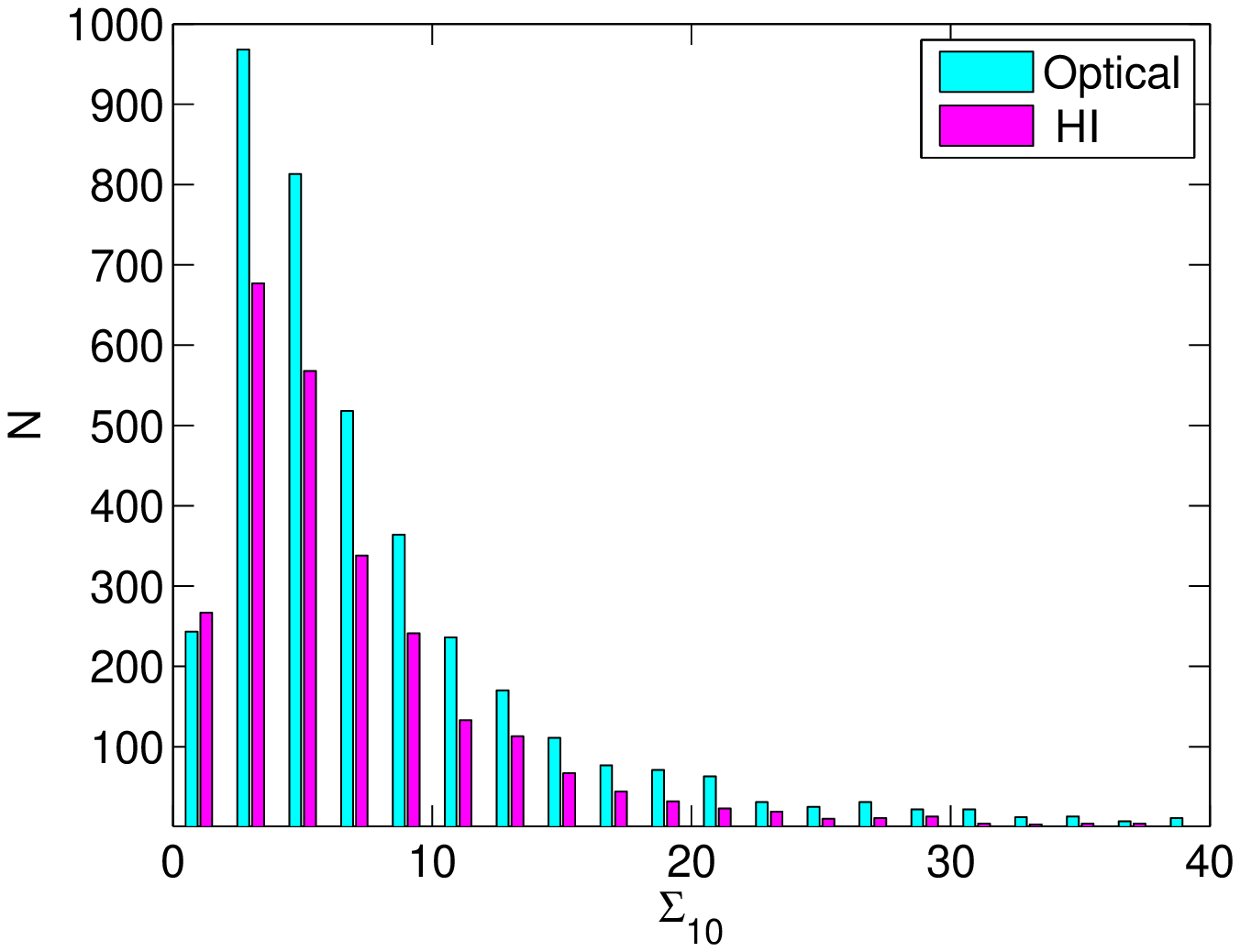,angle=0,width=3.5in}
\psfig{file=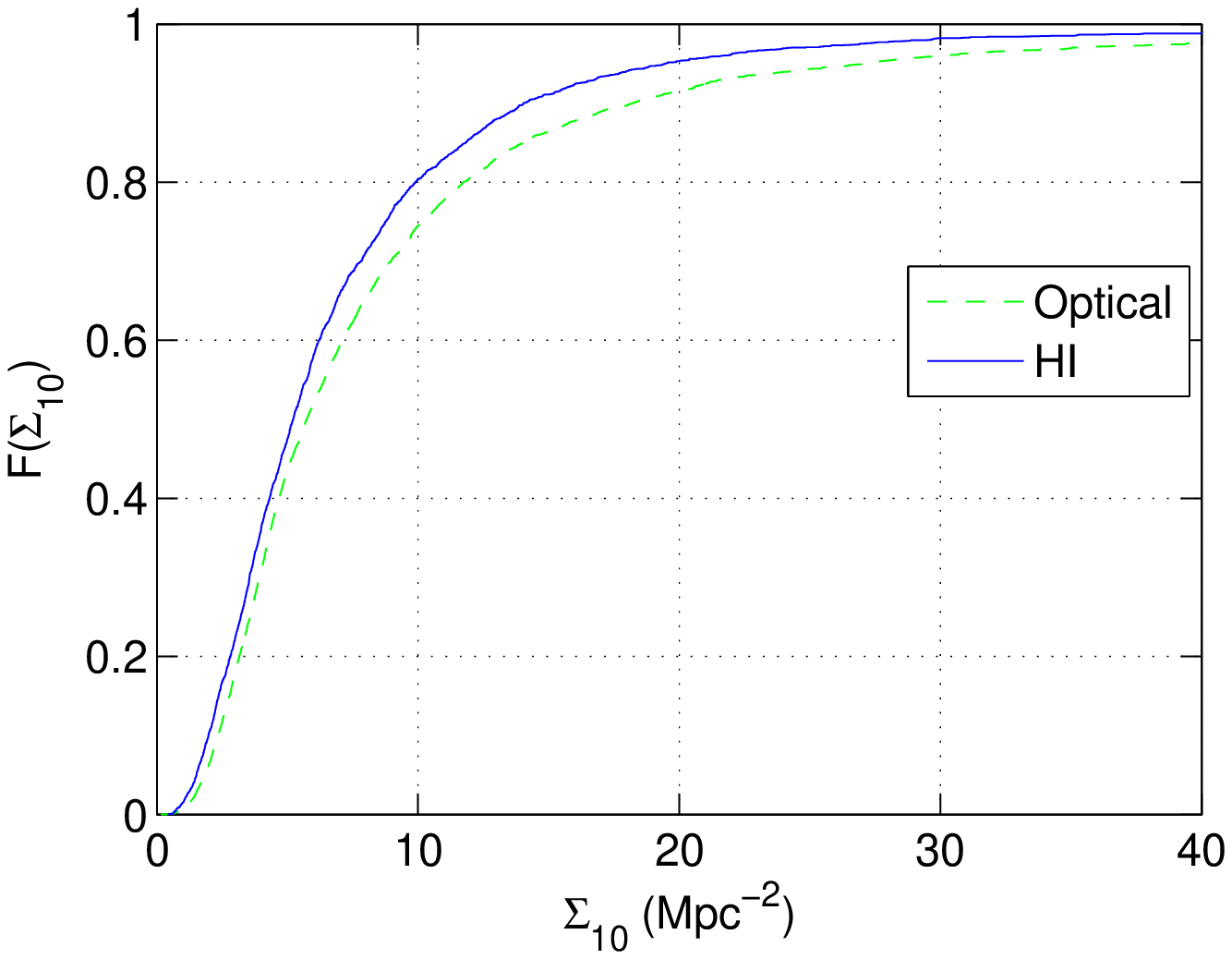,angle=0,width=3.5in}}
  \caption{\small{HOPCAT and 6dFGSDR2 $\Sigma_{10}$ calculated using 6dF2MASS as the background catalogue. Left Panel: Direct comparison of the number of HOPCAT and 6dFGSDR2 galaxies with respect to $\Sigma_{10}$}. Right panel: K-S test showing the two distributions are not the same. Note at $\Sigma_{10}$=5 gal Mpc$^{-2}$ there is an increase in the difference between the two distributions, indicating an increase in the number of optical galaxies. Note the $\Sigma_{10}$ horizontal axis are limited to 40 gal Mpc$^{-2}$}
  \label{Fig:Paper26dFK6dfCompareHopcatK6dFSepHists}
\end{figure*}

The left panel of Fig. \ref{Fig:Paper26dFK6dfCompareHopcatK6dFSepHists} illustrates the difference between HI and optically detected galaxies with respect to $\Sigma_{10}$. Though the 6dFGSDR2 sample (3901 galaxies) is larger than the HOPCAT target list (2602 galaxies), there is a larger number of HI than optical galaxies for $\Sigma_{10}$ less than 1 gal Mpc$^{-2}$.  On running a  Kolmogorov-Smirnov (K-S) test, (right panel Fig. \ref{Fig:Paper26dFK6dfCompareHopcatK6dFSepHists}), we find the null hypothesis can be rejected with a significance level of 0.1 per cent, therefore the two distributions are not the same. At $\Sigma_{10}$=5 gal Mpc$^{-2}$ the difference between the two distributions starts to increase.  This indicates an increase in the number of optically galaxies compared to HI galaxies.  We conclude, as $\Sigma_{10}$ increases, the number of HI galaxies decreases and the number of optically detected galaxies increases.

\subsection{SFR-$\Sigma_{10}$ Relationship \label{SubSec:Paper2ResultsSFRSigma10}}

We use the two HOPCAT target galaxy lists, as described in Sections \ref{SubSec:Paper2IRAS} and \ref{SubSec:Paper2SumssSN6ICats} and Table \ref{Tab:Paper2DataSetNames} to investigate the relationship between SFR and $\Sigma_{10}$ for HI galaxies.

\begin{figure*}
  \centerline{\psfig{file=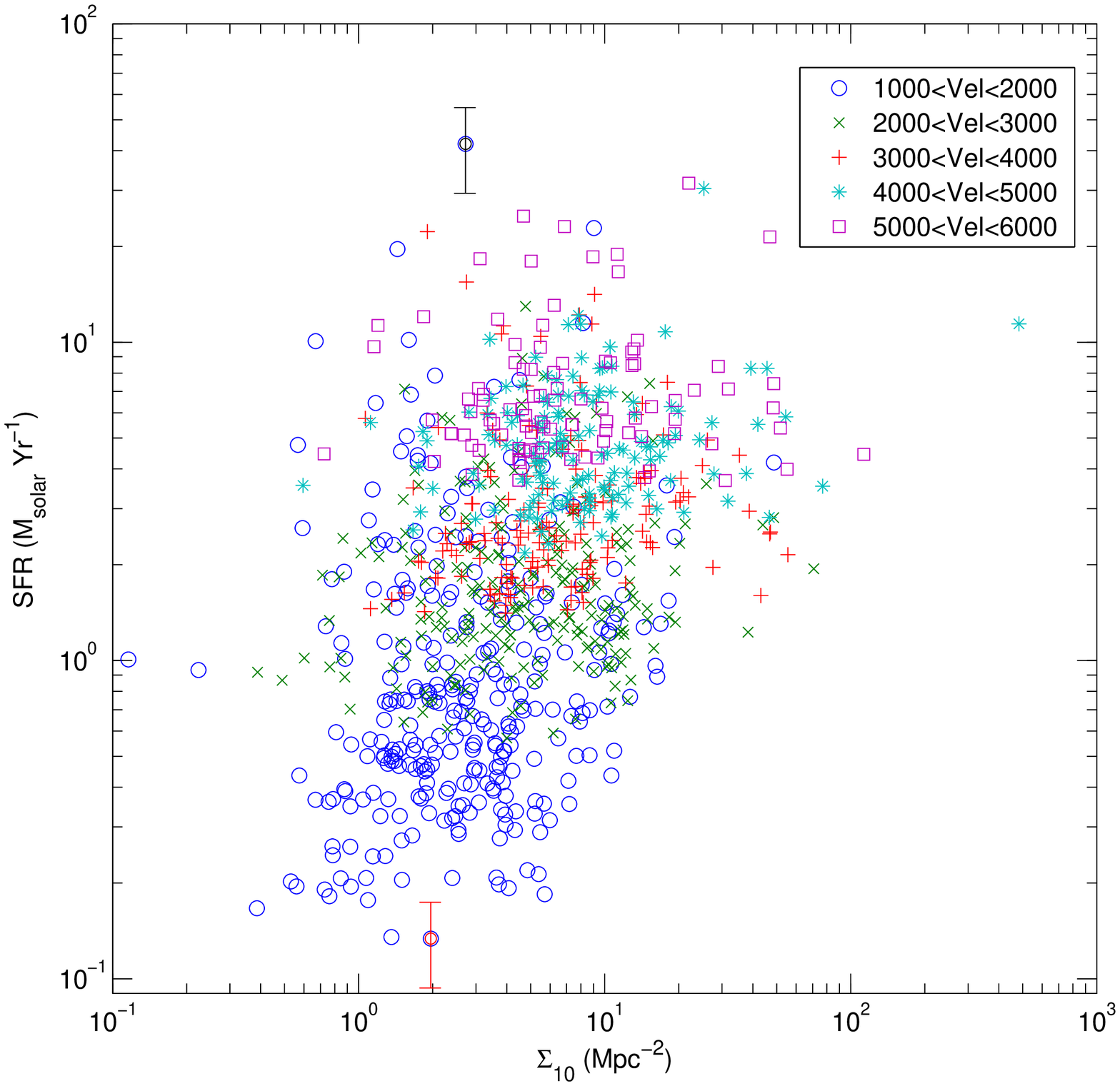,angle=0,width=3.5in}
\psfig{file=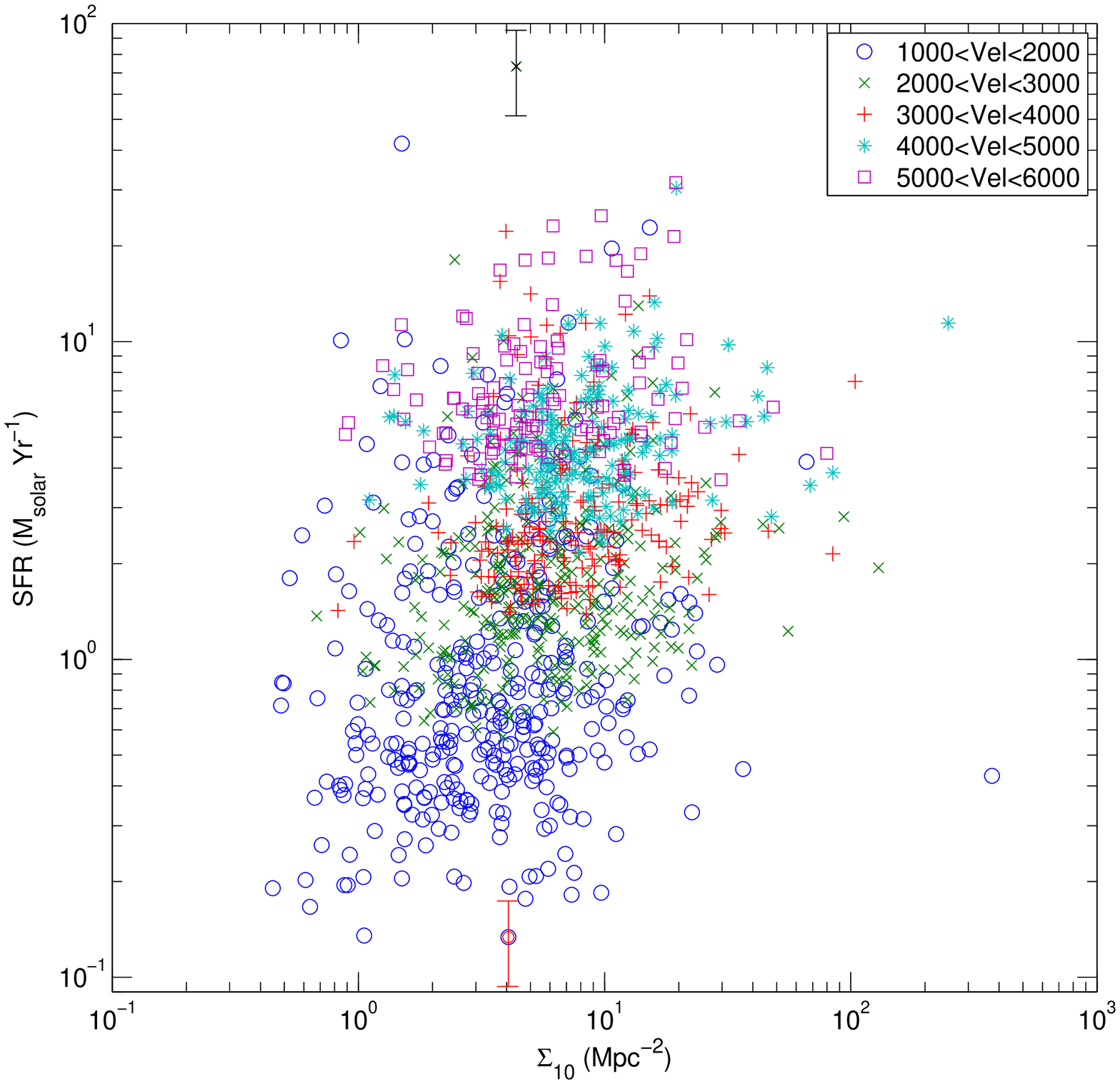,angle=0,width=3.5in} }

\centerline{\psfig{file=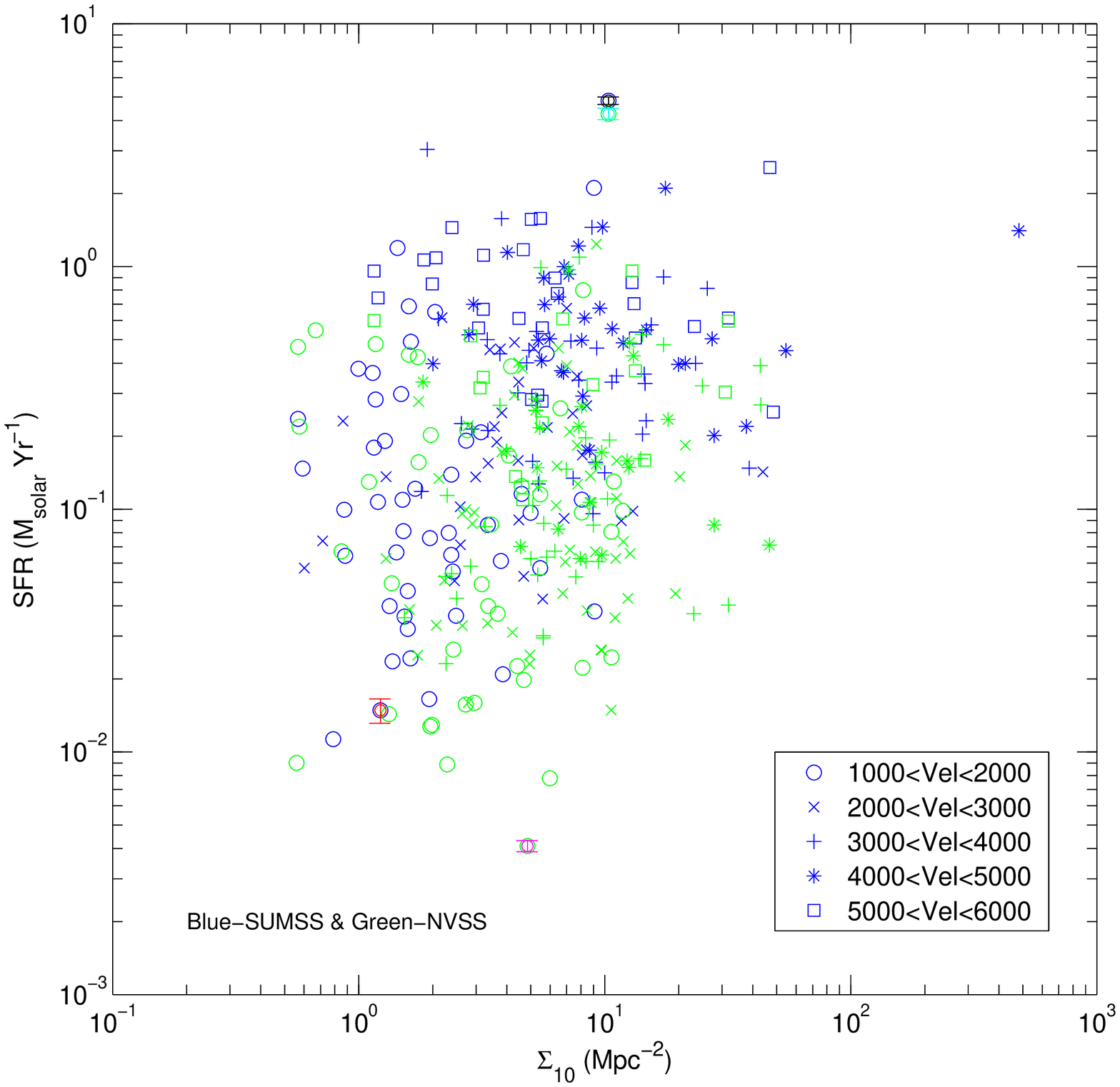,angle=0,width=3.5in}
\psfig{file=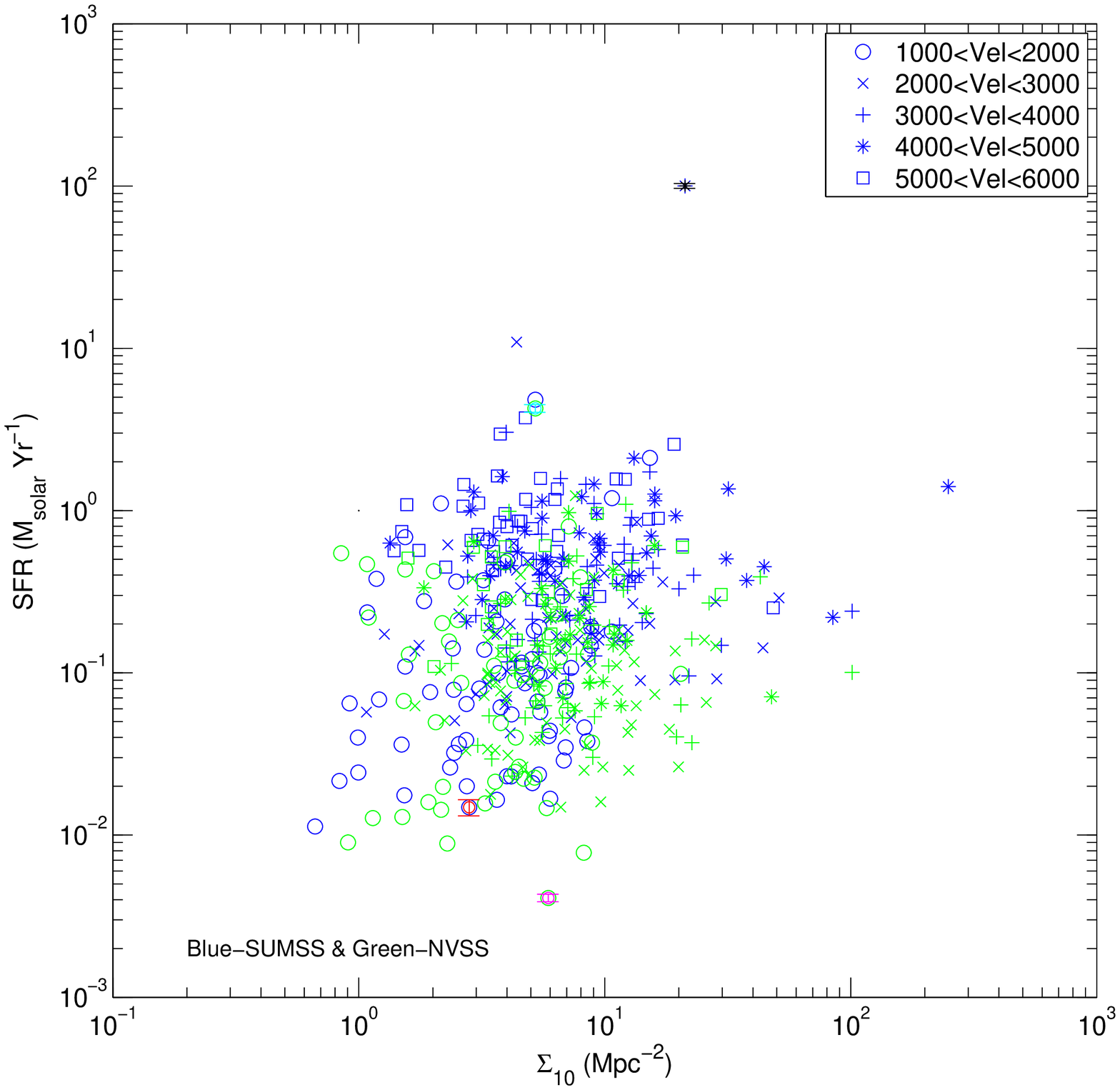,angle=0,width=3.5in}}
  \caption{\small{SFR versus $\Sigma_{10}$. Top Left panel: HOPCAT-SCosGC-IR. Top Right panel: HOPCAT-6dF2MASS-IR. Bottom Left panel: HOPCAT-SCosGC-1.4-GHz. Bottom Right panel: HOPCAT-6dF2MASS-1.4-GHz. The uncertainty values for IR SFRs are shown as minimum (red) and maximum (black). The SUMSS uncertainties are red (min) and black (max) and the NVSS uncertainties are magenta (min) and cyan (max). Due to the multiple line of best fit possible for these plots no conclusion can be drawn.}}
  \label{Fig:Paper2SFRVsSigma10IrasSumssSN6I} 
\end{figure*}

Fig. \ref{Fig:Paper2SFRVsSigma10IrasSumssSN6I} shows the relationship between SFR and $\Sigma_{10}$ for, IR (top panels) and 1.4-GHz (bottom panels) SFRs, for galaxies with 1000$<$vel$<$6000 km s$^{-1}$. Least-squares line of best fit, as described earlier, are fitted. The SFR uncertainty values are shown as red (min) and black (max). The SFR uncertainties for SUMSS are red (min) and black (max) and for NVSS, magenta (min) and cyan (max).

In \citet{gomez2003} and \citet{lewis2002} the average SFR, for a given density range, decreases with increasing $\Sigma_{10}$. We use scatter plots to compare the SFR and $\Sigma_{10}$ for each HI galaxy. In Fig. \ref{Fig:Paper2SFRVsSigma10IrasSumssSN6I} we do not see a clear decrease in the SFR with increasing $\Sigma_{10}$. No conclusions can be drawn from Fig. \ref{Fig:Paper2SFRVsSigma10IrasSumssSN6I} as multiple lines of best fit are possible and the flux-limited nature of HOPCAT maybe affecting the results.

%

\begin{figure*}
  \centerline{\psfig{file=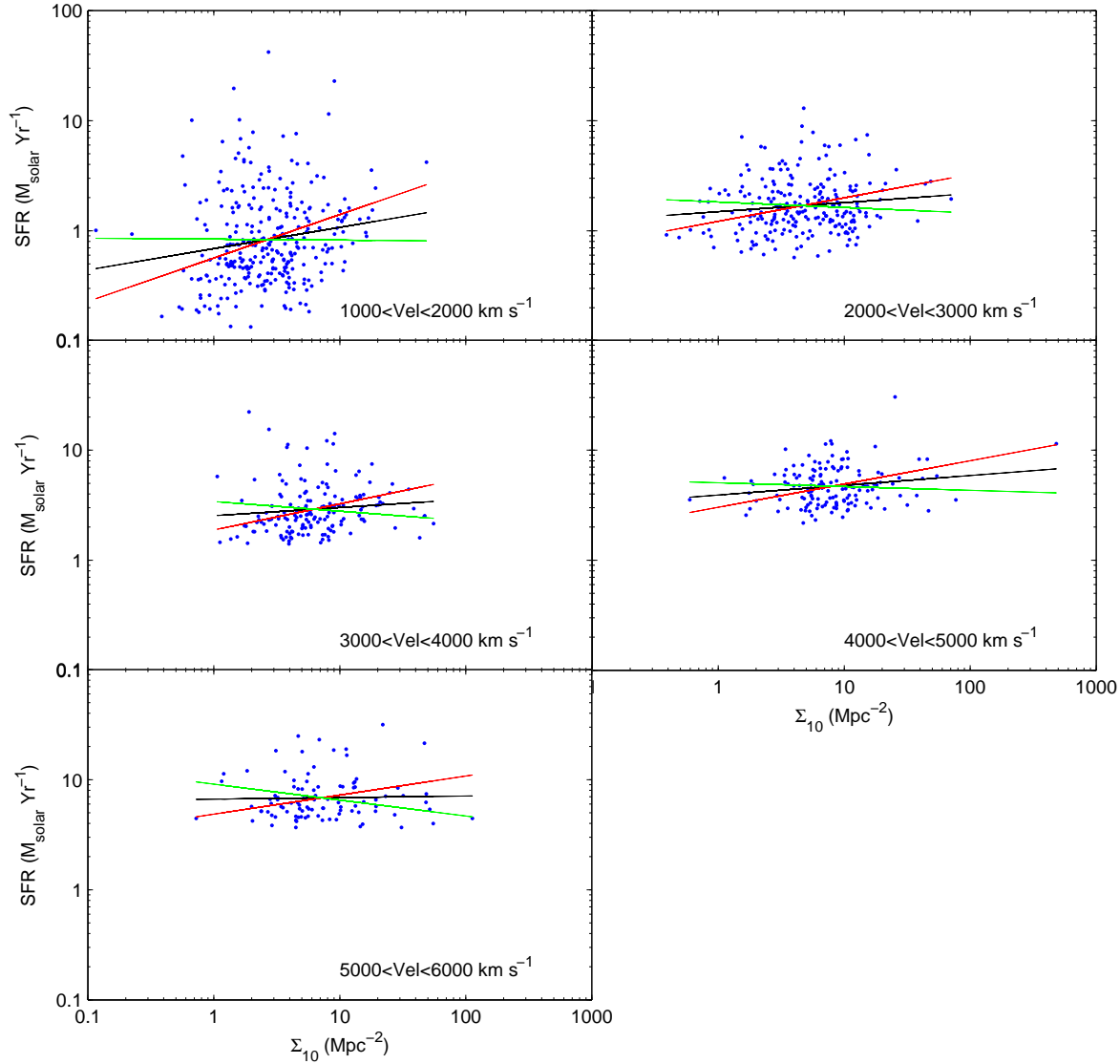,angle=0,width=7in}}
  \caption{\small{SFR versus $\Sigma_{10}$ for HOPCAT-SCosGC-IR, separated into velocity ranges of 1000 km s$^{-1}$. Least-squares line of best fit, calculated using the logs of $\Sigma_{10}$ and SFR. The gradient uncertainty lines are 3$\sigma$.  The SFR remains unchanged with increasing $\Sigma_{10}$.  We compare the distributions of low and high $\Sigma_{10}$ using T-tests and find the null hypothesis cannot be rejected to a 0.1 per cent significance level. See Table \ref{Tab:Paper2TtestResults} for T-test results.
}}
  \label{Fig:Paper2SFRVsSigma10IrasSCosMulti}
\end{figure*}

\begin{figure*}
  \centerline{\psfig{file=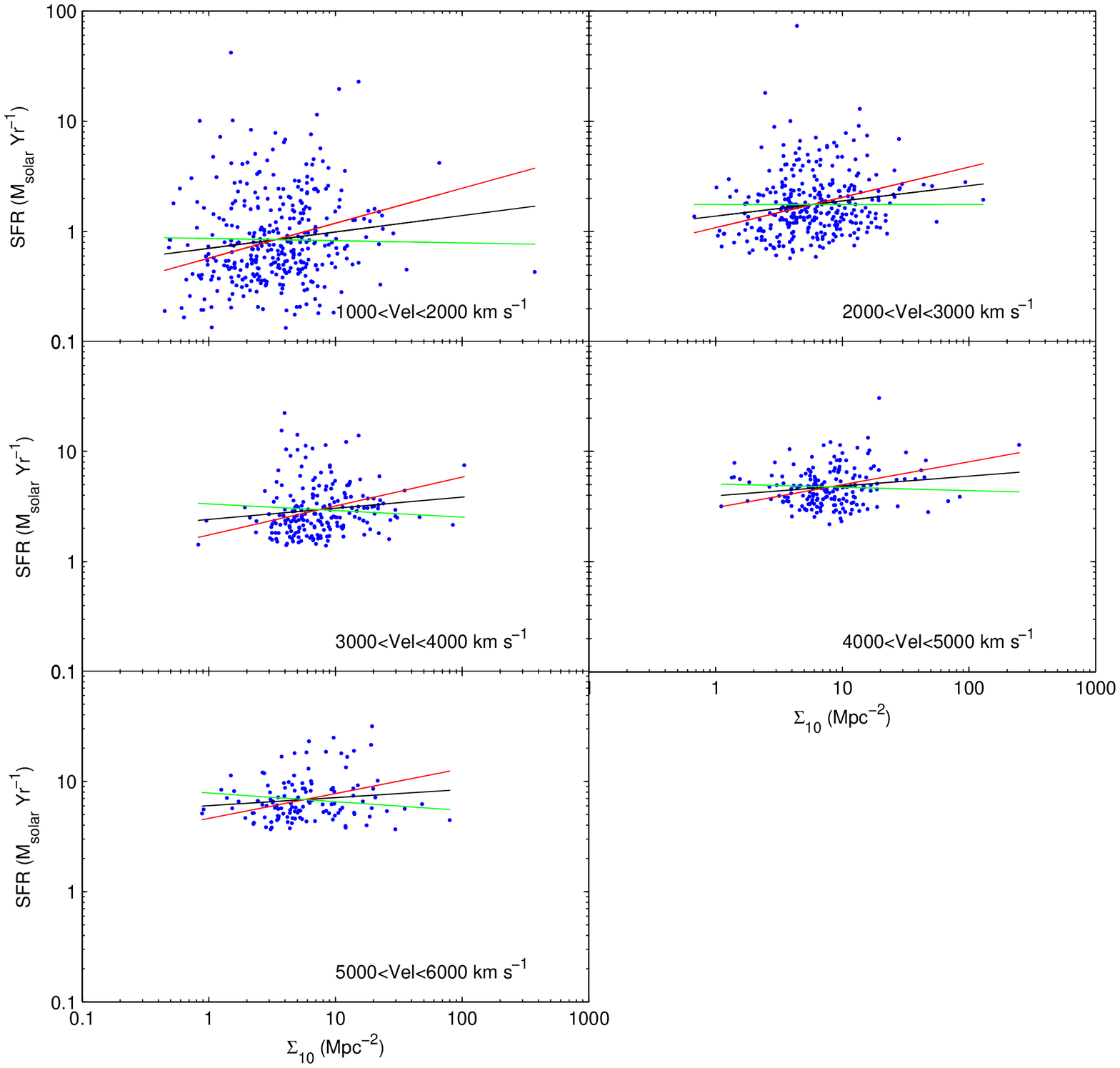,angle=0,width=7in}}
  \caption{\small{SFR versus $\Sigma_{10}$ for HOPCAT-6dF2MASS-IR, separated into velocity ranges of 1000 km s$^{-1}$. Least-squares line of best fit, as described in Fig. \ref{Fig:Paper2SFRVsSigma10IrasSCosMulti}, are fitted. Again the SFR remains unchanged with increasing $\Sigma_{10}$.  We compare the distributions of low and high $\Sigma_{10}$ using T-tests and find the null hypothesis cannot be rejected to a 0.1 per cent significance level. See Table \ref{Tab:Paper2TtestResults} for T-test results.
}}
  \label{Fig:Paper2SFRVsSigma10IrasK6dFMulti}
\end{figure*}

\begin{figure*}
  \centerline{\psfig{file=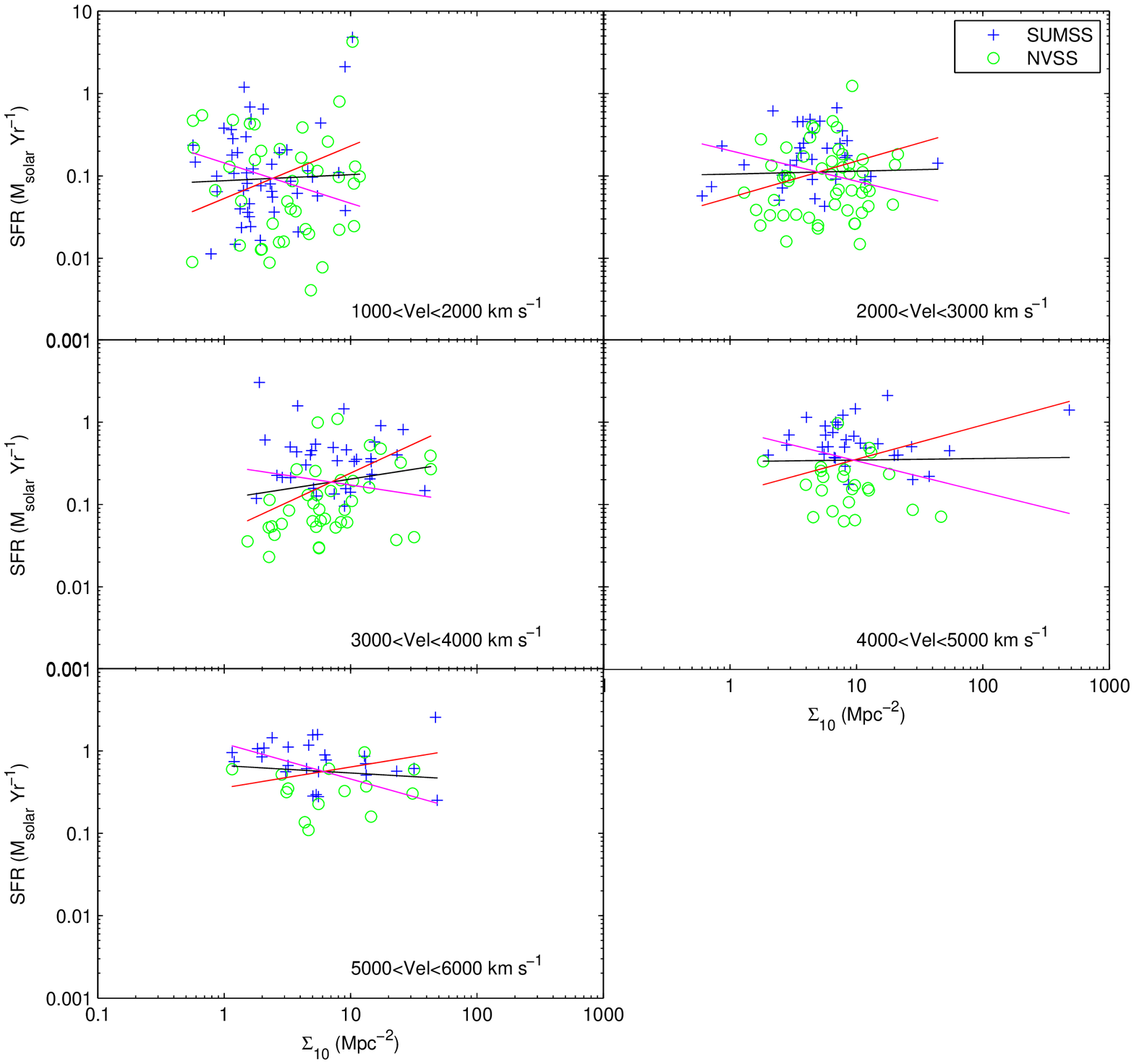,angle=0,width=7in}}
  \caption{\small{SFR versus $\Sigma_{10}$  for HOPCAT-SCosGC-1.4GHz, separated into velocity ranges of 1000 km s$^{-1}$. Least-squares line of best fit, as described in Fig. \ref{Fig:Paper2SFRVsSigma10IrasSCosMulti}, are fitted. The SFR remains unchanged with increasing $\Sigma_{10}$. 
 We compare the distributions of low and high $\Sigma_{10}$ using T-tests and find the null hypothesis cannot be rejected to a 0.1 per cent significance level. See Table \ref{Tab:Paper2TtestResults} for T-test results.
}}
  \label{Fig:Paper2SFRVsSigma10SumssSN6ISCosMulti}
\end{figure*}

\begin{figure*}
  \centerline{\psfig{file=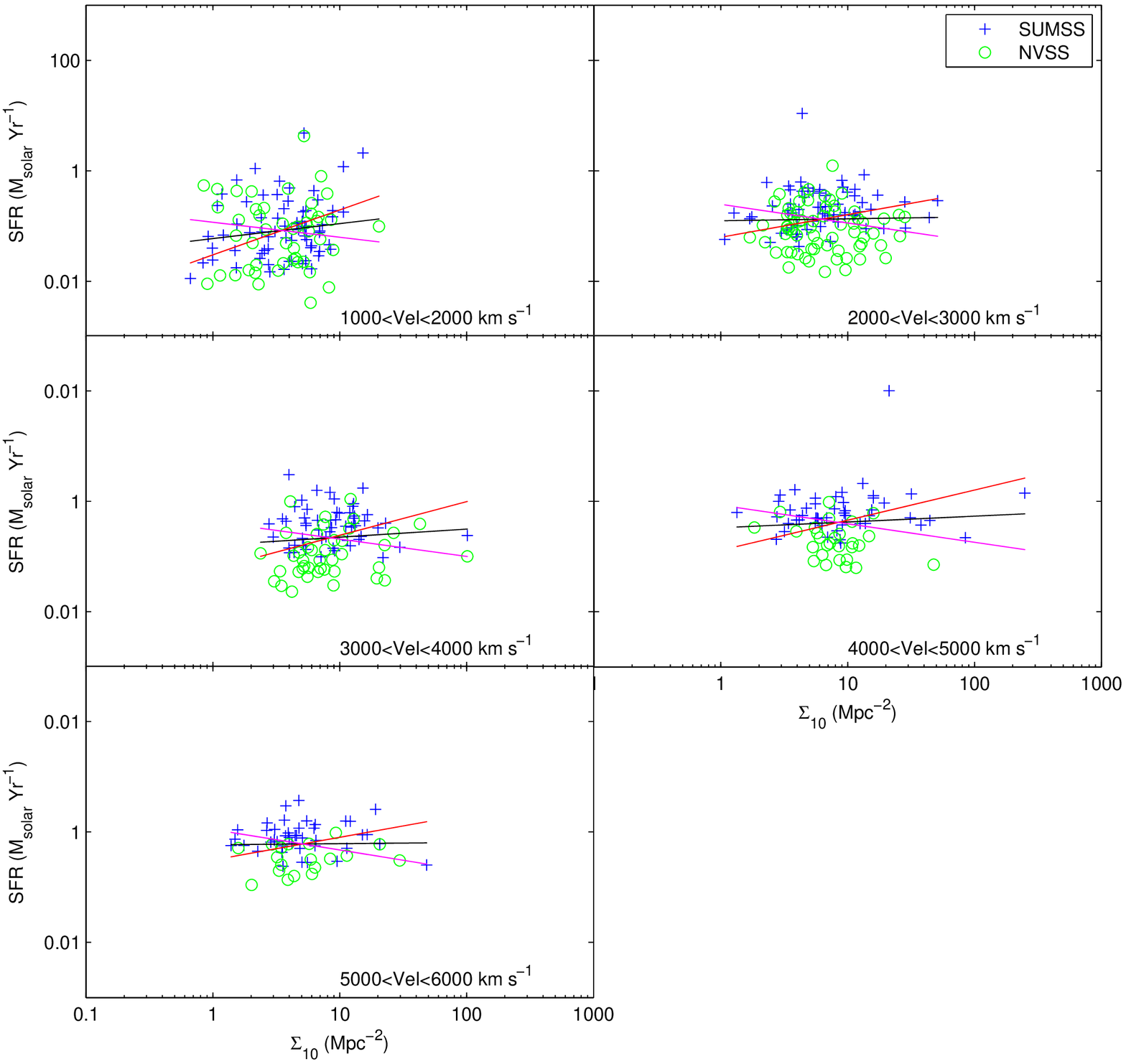,angle=0,width=7in}}
  \caption{\small{SFR versus $\Sigma_{10}$ HOPCAT-6dF2MASS-1.4GHz, separated into velocity ranges of 1000 km s$^{-1}$. Least-squares line of best fit, as described in Fig. \ref{Fig:Paper2SFRVsSigma10IrasSCosMulti}, are fitted.  The SFR remains unchanged with increasing $\Sigma_{10}$.   We compare the distributions of low and high $\Sigma_{10}$ using t-tests and find the null hypothesis cannot be rejected to a 0.1 per cent significance level. See Table \ref{Tab:Paper2TtestResults} for T-test results.
}}
  \label{Fig:Paper2SFRVsSigma10SumssSN6IK6dFMulti}
\end{figure*}

To reduce the flux-limiting effect, we re-analyse SFR versus $\Sigma_{10}$, separated into velocity ranges of 1000 km s$^{-1}$.  In Figs. \ref{Fig:Paper2SFRVsSigma10IrasSCosMulti} and \ref{Fig:Paper2SFRVsSigma10IrasK6dFMulti} (IR) and Figs. \ref{Fig:Paper2SFRVsSigma10SumssSN6ISCosMulti} and \ref{Fig:Paper2SFRVsSigma10SumssSN6IK6dFMulti} (1.4-GHz), the SFR per HI galaxy with respect to $\Sigma_{10}$ remains remarkably constant. The SFR versus $\Sigma_{10}$ average gradients are; for IR 0.11$\pm$0.05, 0.09$\pm$0.15 (Figs. \ref{Fig:Paper2SFRVsSigma10IrasSCosMulti} and \ref{Fig:Paper2SFRVsSigma10IrasK6dFMulti}) and 1.4-GHz 0.12$\pm$0.15 and 0.06$\pm$0.15 (Figs. \ref{Fig:Paper2SFRVsSigma10SumssSN6ISCosMulti} and \ref{Fig:Paper2SFRVsSigma10SumssSN6IK6dFMulti}).

To determine if the relationship is indeed constant, we conduct T-tests, where two normally distributed populations are equal if the null hypothesis cannot be rejected at a low significance level. From the 4 data sets separated into 5 velocity groups, 18 out of a total of 20 the null hypothesis cannot be rejected to a 0.1 per cent significance level. We conclude that the SFR for HI galaxies between 1000 and 6000 km s$^{-1}$ remains constant regardless of the local galaxy density. 

\begin{table*}
\begin{center}
\medskip
\caption{\small{T-test results for SFR and SFE versus $\Sigma_{10}$. \label{Tab:Paper2TtestResults} 
}}
\begin{tabular}{ccccc}
\hline
Velocity  & HOPCAT-  & HOPCAT-  & HOPCAT- & HOPCAT-\\  
Groupings & SCosGC-IR  & 6dF2MASS-IR  & SCosGS-1.4GHz & 6dF2MASS-1.4GHz\\     
km s$^{-1}$ & P-values & P-values & P-values & P-values \\
\hline 
SFR vs $\Sigma_{10}$&Fig. \ref{Fig:Paper2SFRVsSigma10IrasSCosMulti} & Figs. \ref{Fig:Paper2SFRVsSigma10IrasK6dFMulti}& Figs. \ref{Fig:Paper2SFRVsSigma10SumssSN6ISCosMulti} & Fig. \ref{Fig:Paper2SFRVsSigma10SumssSN6IK6dFMulti}\\
1000$<$vel$<$2000	&	0.4790$^\dagger$	&	0.8710$^\dagger$	&	0.3450$^\dagger$	&	0.1830$^\dagger$	\\
2000$<$vel$<$3000	&	0.0953$^\dagger$	&	0.6740$^\dagger$	&	0.7640$^\dagger$	&	0.4410$^\dagger$	\\
3000$<$vel$<$4000	&	0.7920$^\dagger$	&	0.8730$^\dagger$	&	0.8940$^\dagger$	&	0.4850$^\dagger$	\\
4000$<$vel$<$5000	&	0.1500$^\dagger$	&	0.0045$^\dagger$	&	0.9260$^\dagger$	&	0.3140$^\dagger$	\\
5000$<$vel$<$6000	&	0.3370$^\dagger$	&	0.0198$^\dagger$	&	0.7750$^\dagger$	&	0.4460$^\dagger$	\\
\hline
SFE vs $\Sigma_{10}$&Fig. \ref{Fig:Paper2SFEvsSigma10IrasSCosMulti} & Figs. \ref{Fig:Paper2SFEvsSigma10IrasK6dFMulti}& Figs. \ref{Fig:Paper2SFEvsSigma10SumssNvssSCosMulti} & Fig. \ref{Fig:Paper2SFEvsSigma10SumssNvssK6dFMulti}\\
1000$<$vel$<$2000	&	0.2360$^\dagger$	&	0.9850$^\dagger$	&	0.1990$^\dagger$	&	0.1660$^\dagger$	\\
2000$<$vel$<$3000	&	0.4230$^\dagger$	&	0.3470$^\dagger$	&	0.4260$^\dagger$	&	0.2640$^\dagger$	\\
3000$<$vel$<$4000	&	0.8910$^\dagger$	&	0.9700$^\dagger$	&	0.8960$^\dagger$	&	0.7940$^\dagger$	\\
4000$<$vel$<$5000	&	0.5460$^\dagger$	&	0.3270$^\dagger$	&	0.0070$^\ddagger$	&	0.3420$^\dagger$	\\
5000$<$vel$<$6000	&	0.5580$^\dagger$	&	0.0137$^\dagger$	&	0.5400$^\dagger$	&	0.2010$^\dagger$	\\
\hline
\end{tabular}
\end{center}
\footnotesize
\begin{flushleft}
$^\dagger$: h=0 - two normally distributed populations are equal if the null hypothesis cannot be rejected at a significance level of 0.1\%. \\
$^\ddagger$: h=1 - two normally distributed populations are not equal if the null hypothesis can be rejected at a  significance level of 0.1\%.\\
\end{flushleft}
\normalsize 
\end{table*} 

\subsection{SFE Versus $\Sigma_{10}$ \label{SubSec:Paper2ResultsSFESigma10}}

\begin{figure*}
  \centerline{\psfig{file=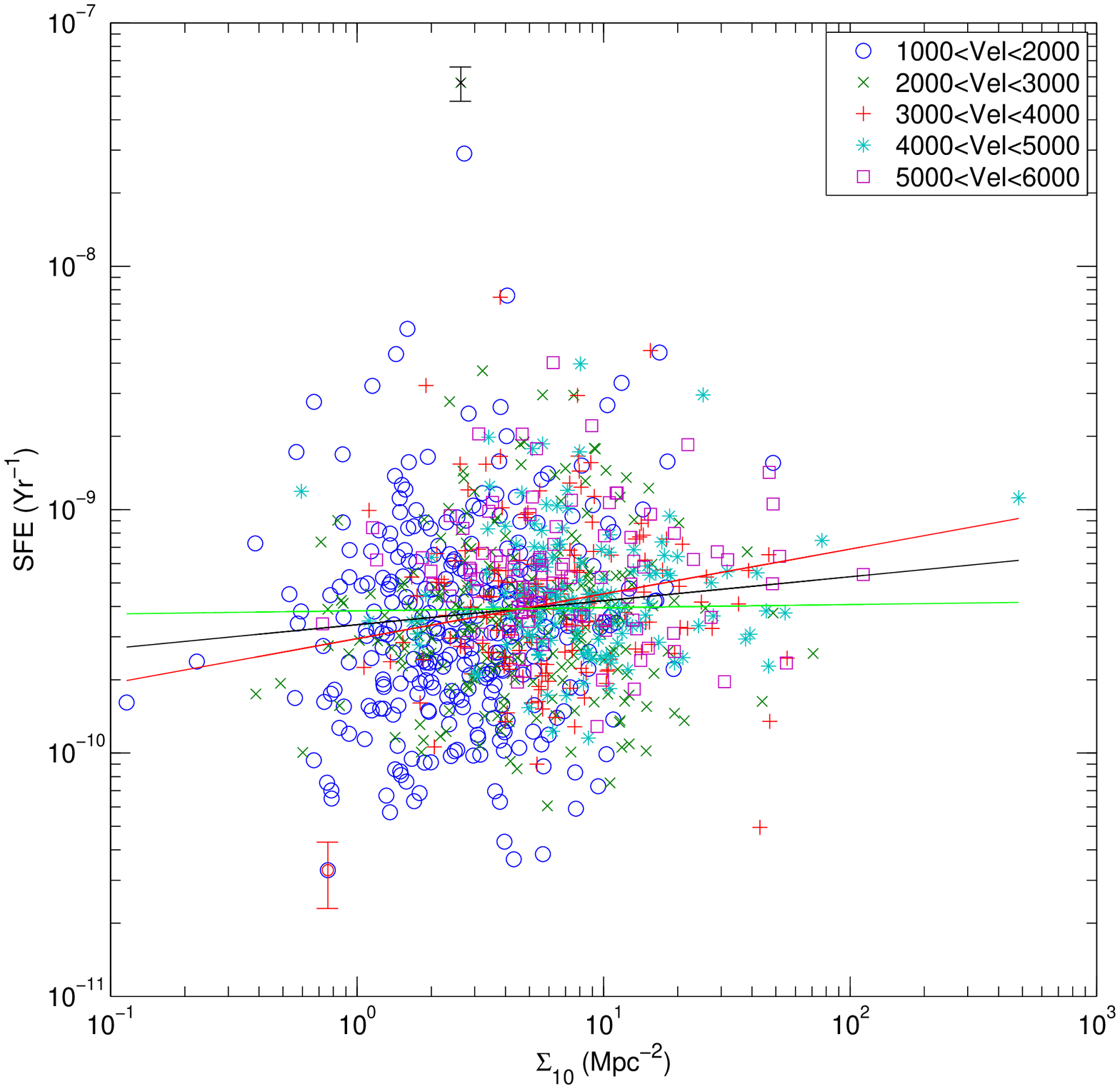,angle=0,width=3.5in}
\psfig{file=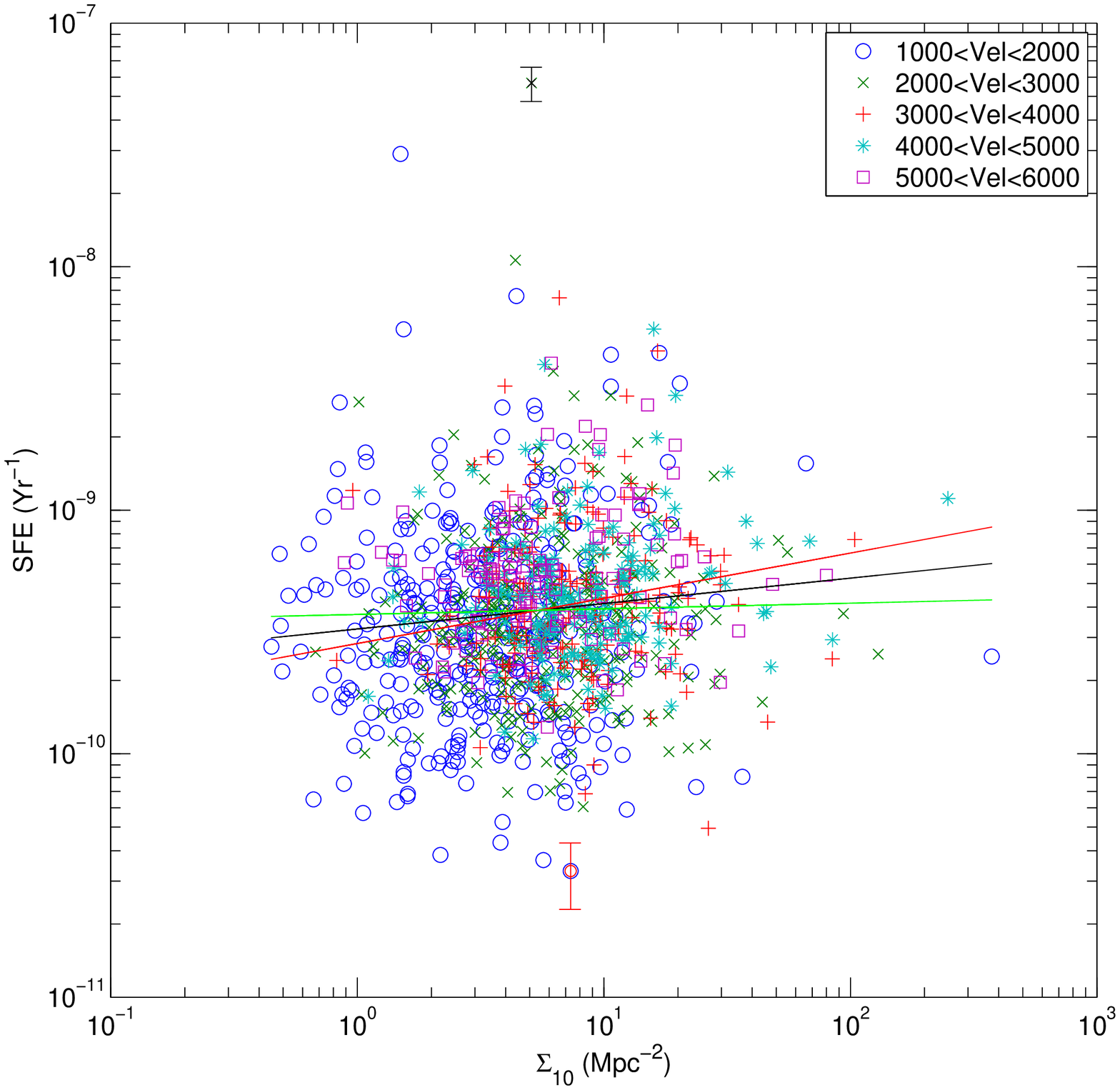,angle=0,width=3.5in} }
  \caption{\small{SFE versus $\Sigma_{10}$. Left panel: HOPCAT-SCosGC-IR. Right panel: HOPCAT-6dF2MASS-IR. Least-squares line of best fit, calculated using the logs of $\Sigma_{10}$ and SFE, with 3$\sigma$ gradient uncertainty lines.  The minimum (red) and maximum (black) uncertainties are shown. The SFE remains unchanged with increasing local galaxy density. 
}}
  \label{Fig:Paper2SFEvsSigma10Iras}
\end{figure*}

\begin{figure*}
  \centerline{\psfig{file=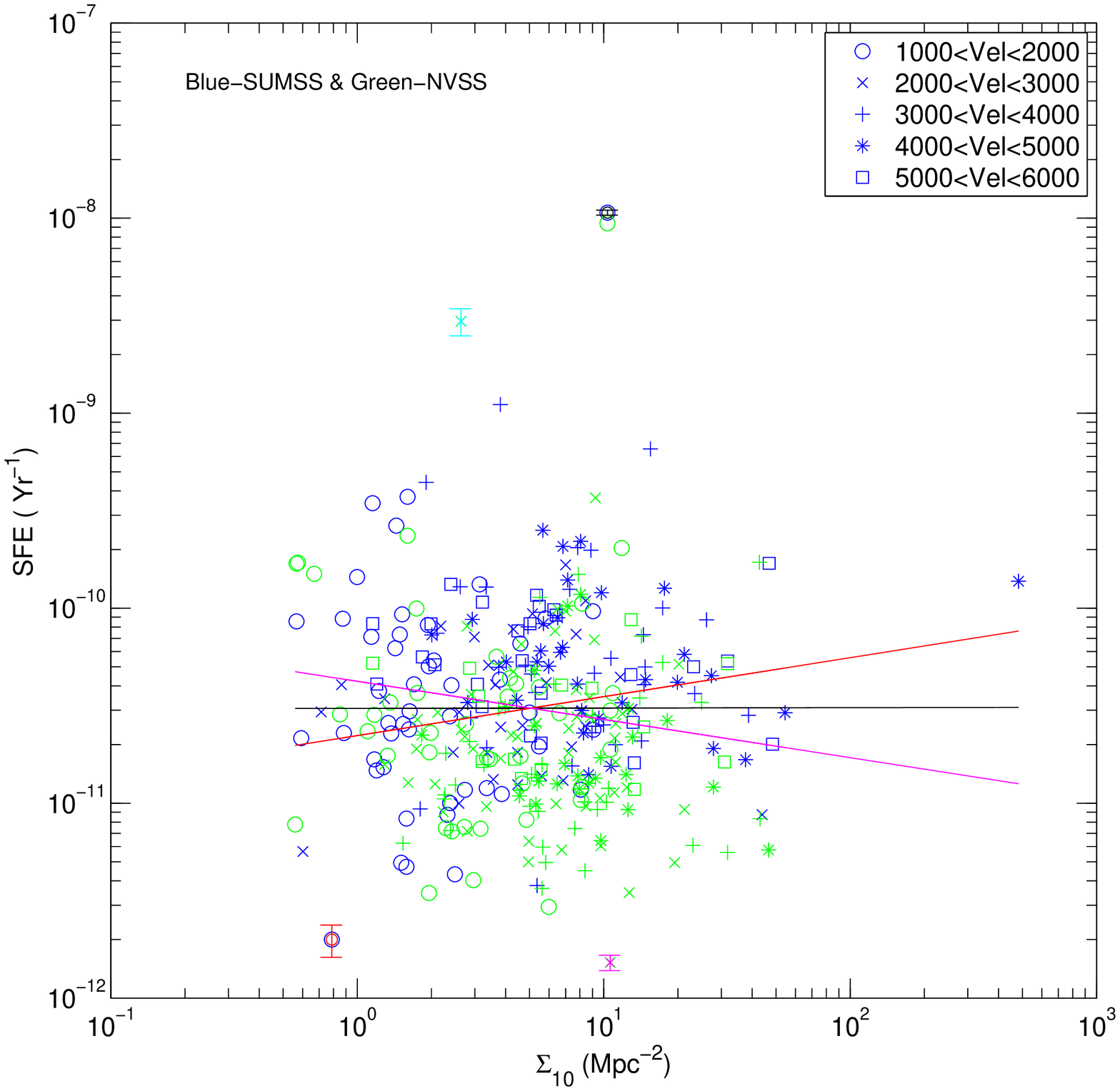,angle=0,width=3.5in}
\psfig{file=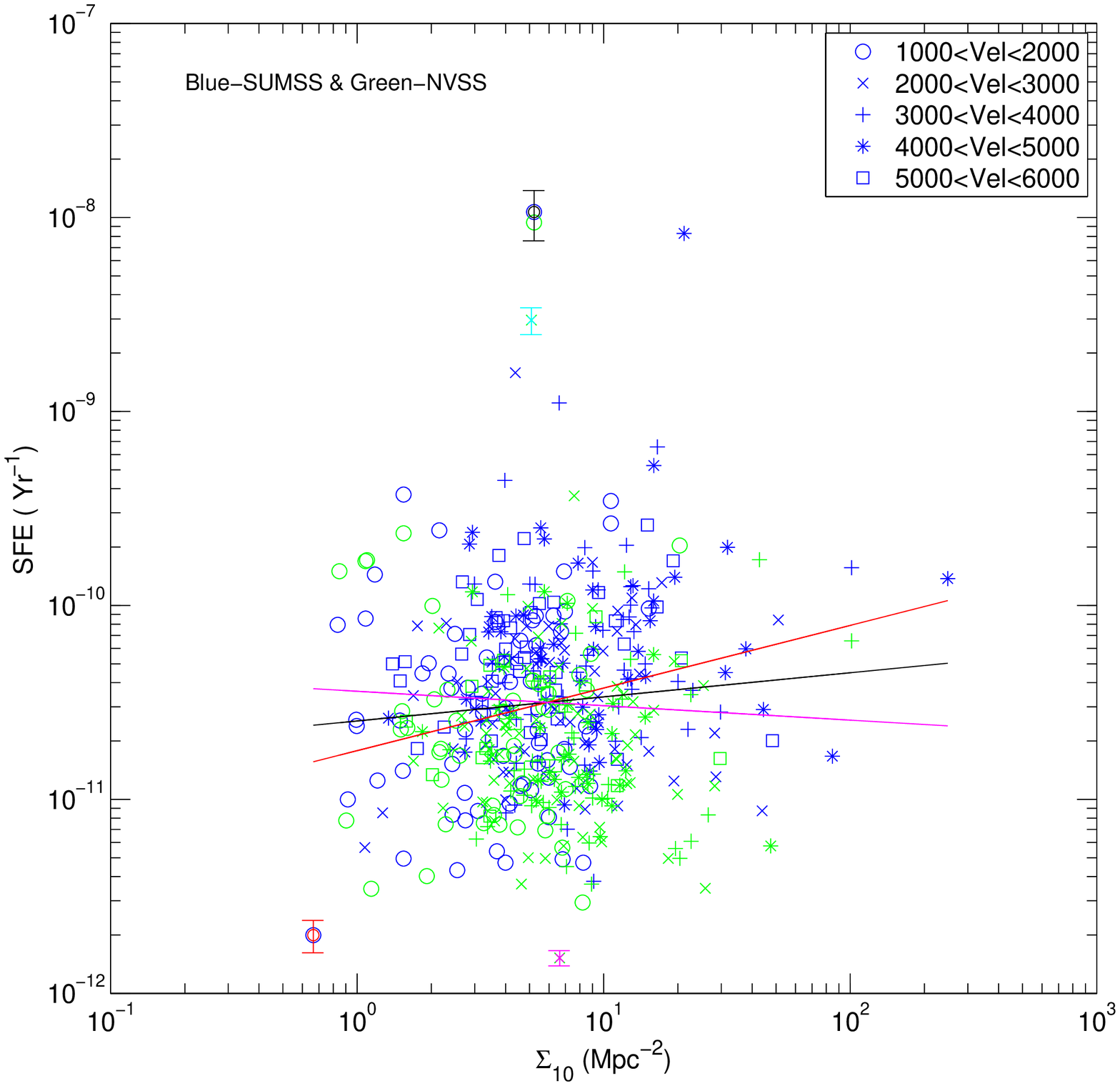,angle=0,width=3.5in}}
  \caption{\small{SFE versus $\Sigma_{10}$. Left panel: HOPCAT-SCosGC-1.4-GHz. Right panel: HOPCAT-6dF2MASS-1.4-GHz. Least-squares line of best fit, as described in Fig. \ref{Fig:Paper2SFEvsSigma10Iras}, are fitted. The SFE remains unchanged with increasing $\Sigma_{10}$. The SFE's uncertainty are shown in red (min) and black (max) for SUMSS and magenta (min) and cyan (max) for NVSS. Like the infrared, the SFE is unchanged with respect to local galaxy density.}}
  \label{Fig:Paper2SFEvsSigma10SumssSN6I}
\end{figure*}

We investigate the efficiency of HI's conversion into stars for local galaxy density using SFE and $\Sigma_{10}$. Fig. \ref{Fig:Paper2SFEvsSigma10Iras} shows the relationship between SFE and $\Sigma_{10}$ for IR and Fig. \ref{Fig:Paper2SFEvsSigma10SumssSN6I} for 1.4-GHz SFEs. Again the sample only contains galaxies with velocities  between 1000 and 6000 km s$^{-1}$. Least-squares line of best fit, as described earlier, are fitted.  The IR $\Delta$SFE's are shown in red (min) and black (max). The 1.4-GHz SFE's uncertainty are shown in red (min) and black (max) for SUMSS and magenta (min) and cyan (max) for NVSS.  For both IR and 1.4-GHz, the SFEs remains unchanged with respect to $\Sigma_{10}$.  

\begin{figure*}
  \centerline{\psfig{file=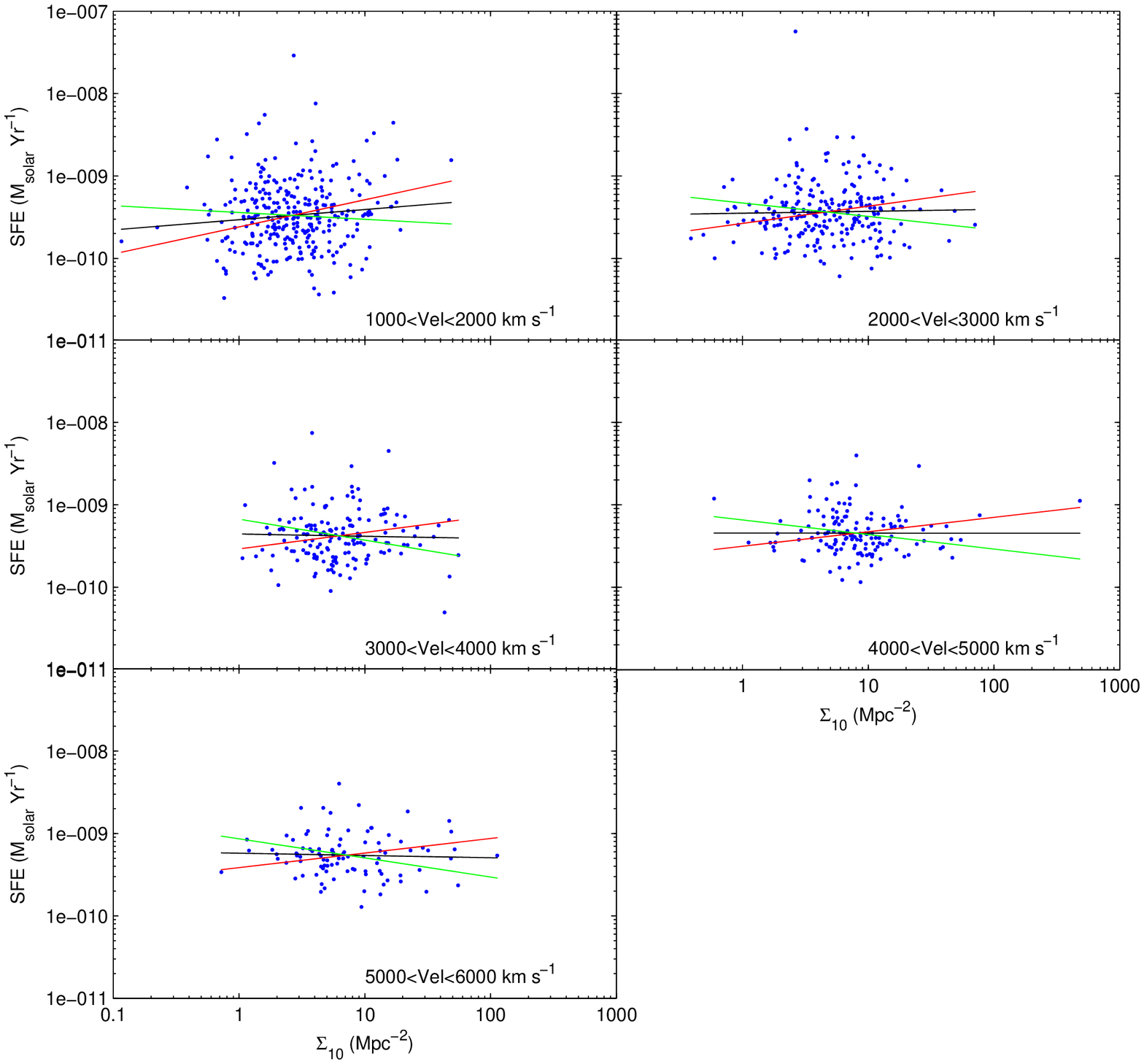,angle=0,width=7in}}
  \caption{\small{SFE versus $\Sigma_{10}$ for HOPCAT-SCosGC-IR, separated into velocity ranges of 1000 km s$^{-1}$. Least-squares line of best fit, as described in Fig. \ref{Fig:Paper2SFEvsSigma10Iras}, are fitted. The SFE remains unchanged with increasing $\Sigma_{10}$.  We compare the distributions of low and high $\Sigma_{10}$ using T-tests and find the null hypothesis cannot be rejected to a 0.1 per cent significance level. See Table \ref{Tab:Paper2TtestResults} for T-test results.}}
  \label{Fig:Paper2SFEvsSigma10IrasSCosMulti}
\end{figure*}

\begin{figure*}
  \centerline{\psfig{file=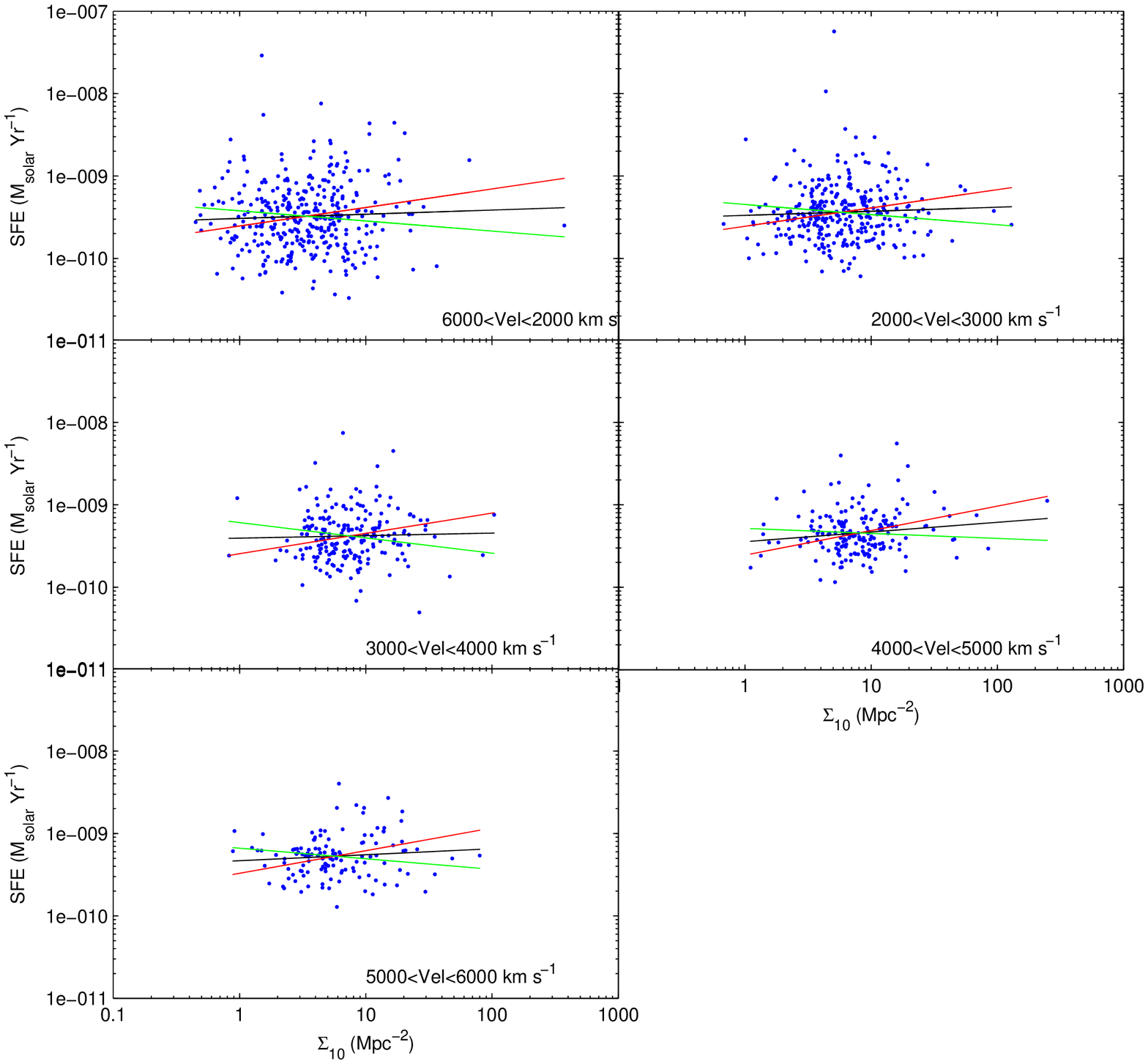,angle=0,width=7in}}
  \caption{\small{SFE versus $\Sigma_{10}$ for HOPCAT-6dF2MASS-IR, separated into velocity ranges of 1000 km s$^{-1}$. Least-squares line of best fit, as described in Fig. \ref{Fig:Paper2SFEvsSigma10Iras}, are fitted. The SFE remains unchanged with increasing $\Sigma_{10}$.  We compare the distributions of low and high $\Sigma_{10}$ using T-tests and find the null hypothesis cannot be rejected to a 0.1 per cent significance level. See Table \ref{Tab:Paper2TtestResults} for T-test results.}}
  \label{Fig:Paper2SFEvsSigma10IrasK6dFMulti}
\end{figure*}

\begin{figure*}
  \centerline{\psfig{file=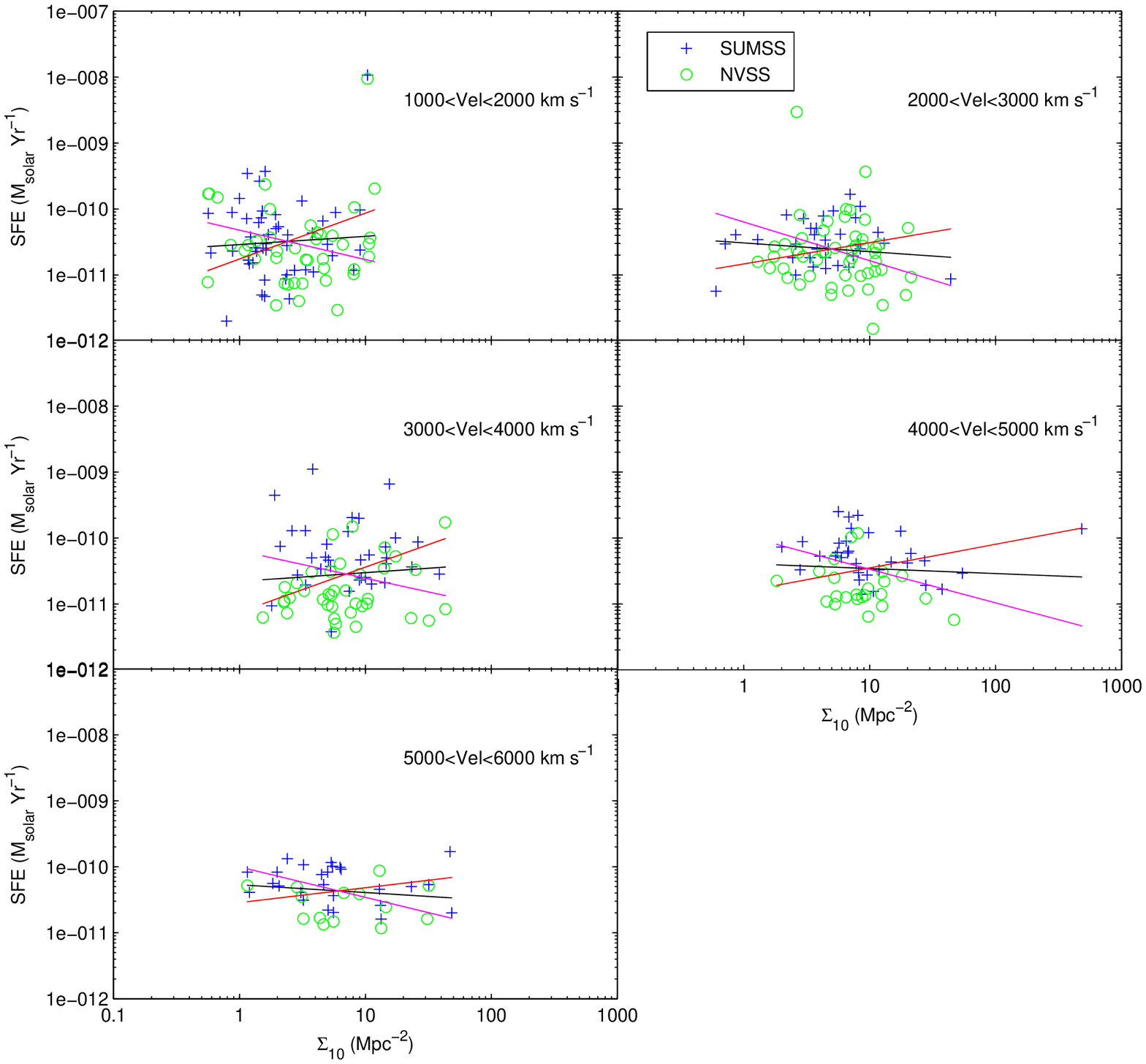,angle=0,width=7in}}
  \caption{\small{SFE versus $\Sigma_{10}$ for HOPCAT-SCosGC-1.4GHz, separated into velocity ranges of 1000 km s$^{-1}$. Least-squares line of best fit, as described in Fig. \ref{Fig:Paper2SFEvsSigma10Iras}, are fitted. The SFE remains unchanged with increasing $\Sigma_{10}$. We compare the distributions of low and high $\Sigma_{10}$ using T-tests and find the null hypothesis cannot be rejected to a 0.1 per cent significance level. See Table \ref{Tab:Paper2TtestResults} for T-test results.}}
  \label{Fig:Paper2SFEvsSigma10SumssNvssSCosMulti}
\end{figure*}

\begin{figure*}
  \centerline{\psfig{file=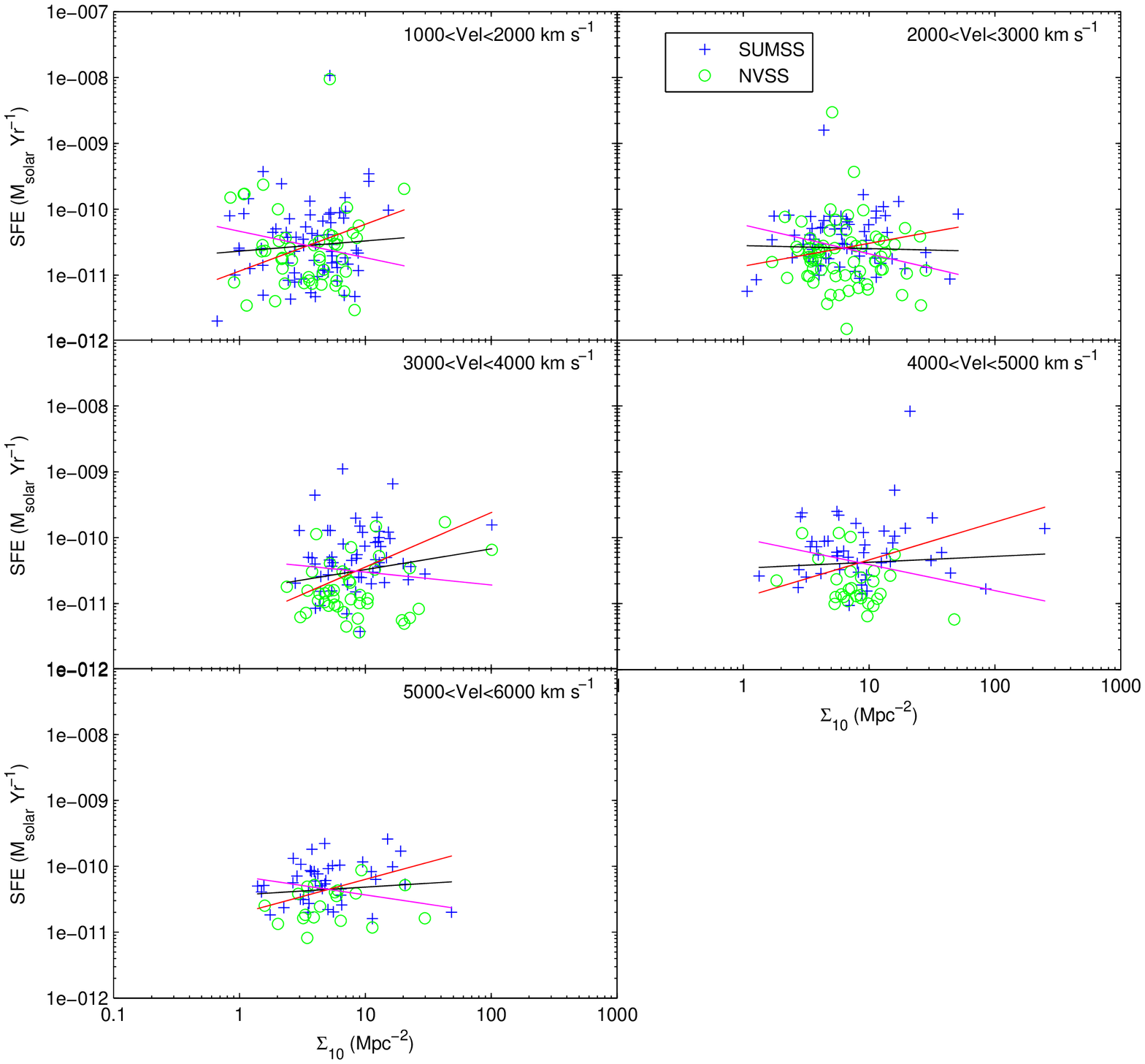,angle=0,width=7in}}
  \caption{\small{SFE versus $\Sigma_{10}$ for HOPCAT-6dF2MASS-1.4GHz, separated into velocity ranges of 1000 km s$^{-1}$. Least-squares line of best fit, as described in Fig. \ref{Fig:Paper2SFEvsSigma10Iras}, are fitted. The SFE remains unchanged with increasing $\Sigma_{10}$.  We compare the distributions of low and high $\Sigma_{10}$ using T-tests and find the null hypothesis cannot be rejected to a 0.1 per cent significance level. See Table \ref{Tab:Paper2TtestResults} for T-test results.}}
  \label{Fig:Paper2SFEvsSigma10SumssNvssK6dFMulti}
\end{figure*}

To ensure the flux-limited nature of our data is not affecting the results, we separate our data into velocity ranges of 1000 km s$^{-1}$. Figs. \ref{Fig:Paper2SFEvsSigma10IrasSCosMulti} and \ref{Fig:Paper2SFEvsSigma10IrasK6dFMulti} shows SFE versus $\Sigma_{10}$ for IR and Figs. \ref{Fig:Paper2SFEvsSigma10SumssNvssSCosMulti} and \ref{Fig:Paper2SFEvsSigma10SumssNvssK6dFMulti} for 1.4-GHz.  The line of best fit along with its 3$\sigma$ gradient uncertainty lines are shown.  The SFE versus $\Sigma_{10}$ average gradients are; IR 0.11$\pm$0.05 and 0.09$\pm$0.15 (Figs. \ref{Fig:Paper2SFEvsSigma10IrasSCosMulti} and \ref{Fig:Paper2SFEvsSigma10IrasK6dFMulti}) and 1.4-GHz  0.12$\pm$0.15 and 0.06$\pm$0.15 (Figs. \ref{Fig:Paper2SFEvsSigma10SumssNvssSCosMulti} and \ref{Fig:Paper2SFEvsSigma10SumssNvssK6dFMulti}).

T-Tests are conducted to determine if the constant SFE for $\Sigma_{10}$ is true. From the 4 data sets separated into 5 velocities groups, 19 out of a total of 20 the null hypothesis cannot be rejected to a 0.1 per cent significance level. 

In Figs. \ref{Fig:Paper2SFEvsSigma10IrasSCosMulti}, \ref{Fig:Paper2SFEvsSigma10IrasK6dFMulti}, \ref{Fig:Paper2SFEvsSigma10SumssNvssSCosMulti} and \ref{Fig:Paper2SFEvsSigma10SumssNvssK6dFMulti}, for a single density, the SFE varies, however as $\Sigma_{10}$ increases the SFE remains unchanged.  Our conclusion is that the efficiency, with which the HI present in galaxies is converted to stars, remains constant regardless of local surface density. 

%

%

%

%

\section{Discussion} \label{Sec:Paper2Discussion}
In this section we discuss the relationships between M$_{HI}$ and SFR, SFR and $\Sigma_{10}$, and SFE and $\Sigma_{10}$. We also compare optical and HI detected galaxy samples. From these results we determine which of our two theories explain the SFR-density relation.

\subsection{HI mass-SFR Relationship \label{SubSec:Paper2DiscussMassHISFR}}

We investigate the relationship between the SFR and M$_{HI}$ HOPCAT galaxies with IR and 1.4-GHz SFRs.  We find M$_{HI}$ increases with increasing SFR.  This result is in general agreement with the Schmidt law, a relationship between a galaxy's gas surface density and the disk average SFR as discussed in \cite{kennicutt1998}.

\citet{mirabel1988}, investigating 92 luminous infrared galaxies find that 88 galaxies have M$_{HI}$ in the range of $5\times10^8$ to $3\times10^{10} M_{\odot}$. They conclude the FIR luminosities have only a loose correlation with the M$_{HI}$. However, our study looks at HI galaxies and calculates SFR from IR and 1.4-GHz fluxes and finds a strong correlation between the SFR and M$_{HI}$.  \cite{stevens2005} also investigated the relationship between M$_{HI}$ and SFR using a volume limited subsample of HICAT, containing 85 FIR and 277 radio (NVSS and SUMSS) based SFRs. \cite{stevens2005} also finds M$_{HI}$ increases with increasing SFR.  

Although we do not know where star formation occurs in a galaxy with respect to the HI detection, the correlation between the M$_{HI}$ and SFR shows HI is an indication of star formation.  In fact, as our SFRs are from IR and 1.4-GHz fluxes, which are indicators of current star formation, the M$_{HI}$ and SFR correlation suggests HI indicates `current' star formation in galaxies.

\subsection{HI radio and Optical galaxy samples \label{SubSec:Paper2DiscussCompareHiOp}}

Our result from investigating the difference between optical and HI detected galaxy samples show there is a much larger population of elliptical galaxies than HI detected galaxies with increasing local surface density. This result agrees with studies into the morphology-density relation (\citet{dressler1980b} etc.), which find for an increasing local galaxy density there is a decrease in the number of spiral galaxies and an increase in the number of elliptical galaxies.  It is expected that this decrease in the number of current star forming HI galaxies with increasing $\Sigma_{10}$ will have a direct effect on the results from any SFRs versus $\Sigma_{10}$ analysis for optical samples.

\subsection{SFR and $\Sigma_{10}$ \label{SubSec:Paper2DiscussSFRSigma10}}

From our investigation into SFR and $\Sigma_{10}$ we conclude the SFRs for our HI galaxies are not affected by increasing $\Sigma_{10}$.

As HOPCAT is based on the Parkes HI All Sky Survey (HIPASS) there may be some concern about the number of low surface brightnes (LSB) galaxies in the sample.  The HIPASS galaxy sample is dominated by late-type galaxies as is expected due to the HI selection (e.g. Helmboldt et al. 2004). As \citet{helmboldt2004} show, HI-selected galaxies tend to have lower luminosity and surface brightness than a comparison sample of optically-selected galaxies. This difference is not so great for our HOPCAT sample because our mean redshift is much greater than the Helmboldt sample. Our median luminosity is $M_B=-18.8$ mag compared to $M_B=-18.5$ mag in their HI-selected sample and  $M_B=-19.1$ mag in their optical sample. Some 15 per cent of their HI-selected galaxies have LSB compared to the optical sample: the fraction of LSB galaxies in HOPCAT will be smaller because of the higher median luminosity.  If we consider the final HOPCAT sample with reliable densities and star formation rates used in our analysis, only 1.3 per cent of the galaxies are fainter than the M$_{B_j}$=-17 mag faint absolute magnitude limit applied in the density calculations. 

In \citet{gomez2003} and \citet{lewis2002} the decrease in the average SFR with an increase in local surface density, for a given density range, is explained by their optical sample.  An optical sample contains both spiral and elliptical galaxies. For areas of increasing local galaxy density, such as compact galaxy groups and poor and rich clusters, the number of spiral galaxies decreases as the number of elliptical galaxies increases.  Studies into the morphology-density relation (\citet{dressler1980b} etc.), show this change in morphology type for increasing local galaxy density. Elliptical galaxies have little or no current star formation, hence the average SFR, as \citet{gomez2003} and \citet{lewis2002} use, will decrease in areas of increasing local surface density where elliptical galaxy numbers increase and spiral galaxy numbers decrease.  As our sample does not have elliptical galaxies and the SFRs are for galaxies with current star formation, a more direct picture of SFR with respect to local surface density is seen.  We clearly show the HI galaxies' SFRs are relatively constant regardless of local surface density.  

\citet{gomez2003} and \citet{lewis2002} describe a characteristic $\Sigma_{10}$ break at 1 gal Mpc$^{-2}$ where the average SFR decreases sharply. As our scatter plots are a direct comparison of SFR per HI galaxy for each galaxies local surface density, using lines of best fit, no break is seen.  However, at 5 gal Mpc$^{-2}$ (right panel of Fig. \ref{Fig:Paper26dFK6dfCompareHopcatK6dFSepHists}), the increasing gap indicates a decrease in the number of HI galaxies relative to optical galaxies.  \citet{lewis2002} and \citet{gomez2003} use a faint absolute magnitude limit of -19 mag and we use -17 mag which results in our larger $\Sigma_{10}$ values. Our decrease in HI galaxy numbers at $\Sigma_{10}$=5 gal Mpc$^{-2}$ may coincide with \citet{gomez2003} and \citet{lewis2002} characteristic $\Sigma_{10}$ break at 1 gal Mpc$^{-2}$.

\citet{stevens2005}, also using HICAT, finds the SFR for galaxies in groups is smaller than field galaxies. This differs from the results presented in this paper for the same sample. However, \cite{stevens2005} uses a group finder algorithm to find HI galaxies in groups (143) and compares their SFR to those for field galaxies (17). As the resulting density is 3D and the analysis \cite{stevens2005} uses is significantly different from ours, no direct comparison is possible. 

However, one observation \cite{stevens2005} makes is there is a larger population of red galaxies in groups than in the field. He concludes this could partially explain the lower SFR in group galaxies, which agree with our results.  It is clear the lower average SFR in regions of high local surface density in \citet{lewis2002} and \citet{gomez2003} can be attributed to their larger population of elliptical galaxies.

\citet{balogh2004} using H$\alpha$ equivalent widths to investigate the population of star-forming galaxies in SDSS and 2dFGRS and finds `the distribution of line strength' for these star-forming galaxies is independent of environment.  This appears to be in agreement with our results.

Our conclusion still remains, for HI galaxies between 1000 and 6000 km s$^{-1}$ in the southern sky, the SFR does not depend on the local galaxy density. 

\subsection{Fewer Star Forming Galaxies or Suppressed Star Formation?}

In Section \ref{SubSec:Paper2DiscussMassHISFR} we show a strong correlation exists between M$_{HI}$ and SFR leading to the conclusion our HI galaxies display current star formation. As discussed in Section \ref{SubSec:Paper2DiscussCompareHiOp}, the number of HI galaxies decreases sharply in regions of high local surface density. The difference between this work and \citet{gomez2003} and \citet{lewis2002} is our sample contains current star forming galaxies and theirs are optical galaxy samples which contain galaxies with little or no current star formation.  We also find in Section \ref{SubSec:Paper2DiscussSFRSigma10}, the SFRs do not depend on $\Sigma_{10}$. Our results in Section \ref{SubSec:Paper2ResultsSFESigma10} show the efficiency, with which HI is converted to stars, is not affected by an increase in the local surface density.  We now conclude which theory explains the SFR-density relation.

We find there are, indeed, fewer star forming (HI) galaxies in regions with high local surface density, which affects the average SFR for optical galaxy samples. As the SFR and SFE do not depend on the local surface density, the SFR-density relation is due to the low number of star forming (HI) galaxies, in regions of high local surface density and is not due to the suppression of star formation in these regions.

\section{Summary} \label{Sec:Paper2Summary}
We use an optically matched HI detected galaxy catalogue, HOPCAT, to investigate SFR, SFE and the local surface density to explain the SFR-density relation. In regions of high local galaxy density either, there are fewer star forming galaxies, or the galaxies capable of forming stars are present but some physical process is suppressing their star formation. The local surface density, $\Sigma_{10}$, is calculated using HOPCAT with SCosGS and 6dF2MASS as background catalogues. {\it{IRAS}}, SUMSS and NVSS fluxes are used to calculate the SFRs and SFEs. For the SFR-M$_{HI}$ investigations we use 2 different data sets using 6dF2MASS with IR and 1.4-GHz SFRs, HOPCAT-6dF2MASS-IR (1405 galaxies) and HOPCAT-6dF2MASS-1.4GHz (561 galaxies). For the SFR-SFE-$\Sigma_{10}$ investigations we have 4 different data sets using SCosGS and 6dF2MASS each with IR and 1.4-GHz SFRs, HOPCAT-SCosGS-IR (923 galaxies), HOPCAT-6dF2MASS-IR (1180 galaxies), HOPCAT-SCosGS-1.4GHz (339 galaxies) and HOPCAT-6dF2MASS-1.4GHz (478 galaxies). We find:

1. A strong correlation exists between M$_{HI}$ and SFR for galaxies out to 10000 km s$^{-1}$ in the southern sky. The SFRs are calculated using IR and 1.4-GHz fluxes, both of which are indicators of current star formation.  We conclude HI galaxies display current star formation and their M$_{HI}$ and SFR are strongly related.

2. The number of HI galaxies decreases with increasing local surface density. This agrees with \citet{dressler1980b}, \citet{hashimoto1998}, \citet{martinez2002}, \citet{goto2003}, \citet{sodre2003}, \citet{nuijten2005} and \citet{christlein2005}, where optical samples show an increase in the number of non-current star forming elliptical galaxies with a decrease in current star forming galaxy numbers for increasing local galaxy density. This is also seen in \citet{balogh2004}.

3. For our HI galaxies, the SFR does not depend on local surface density.  Any decrease in the average SFR with increasing local surface density for optical galaxy samples, is explained by the decrease in the number of current star forming galaxies and an increase in elliptical galaxies in regions with increasing local surface density.

4. The efficiency with which the HI present in the HI galaxies is converted into stars is the same regardless of local surface density. 

5. The SFR-density relation is due to a decrease in the number of star forming (HI) galaxies in regions of high local (surface) galaxy density and not due to the suppression of star formation in these regions.

\section*{Acknowledgments}
We thank Elaine Sadler for her assistance with the SUMSS catalogue, completeness and reliability measurements and for her invaluable help and encouragement.  We thank Tom Mauch for the use of his SN6I catalogue and Mike Read for providing the SuperCOSMOS Galaxy Catalogue.  Thanks also to Jamie Stevens, Martin Zwaan and Kevin Pimbblet for their helpful comments. Thanks to the Astrophysics Group at the University of Queensland for their assistance. A very special thanks to Luke Pegg and Katrina Seet.

We acknowledge the assistance the 6dF Galaxy Survey conducted primarily by The Anglo-Australian Observatory, and with the support from the ANU and the Wide-field Astronomy Unit of the University of Edinburgh.  We also acknowledge the use of the {\it{IRAS}} Galaxy and Quasar Catalogue provided by the NASA/ IPAC Infrared Science Archive, which is operated by the Jet Propulsion Laboratory, California Institute of Technology, under contract with the National Aeronautics and Space Administration.

MTD is supported through a University of Queensland Graduate School Scholarship. This work is supported by a University of Queensland Research Development Grant and by DP and LIEF grants from the Australian Research Council.

\bibliographystyle{mn2e}
\bibliography{MarzBib}

\begin{thebibliography}{}

\bibitem[\protect\citeauthoryear{{Allam}, {Tucker}, {Lin} \&
  {Hashimoto}}{{Allam} et~al.}{1999}]{allam1999}
{Allam} S.~S.,  {Tucker} D.~L.,  {Lin} H.,    {Hashimoto} Y.,  1999, ApJL, 522,
  L89

\bibitem[\protect\citeauthoryear{{Balogh} et~al.,}{{Balogh}
  et~al.}{2004}]{balogh2004}
{Balogh} M.,  et~al., 2004, \mnras, 348, 1355

\bibitem[\protect\citeauthoryear{{Cardiel} et~al.,}{{Cardiel}
  et~al.}{2003}]{cardiel2003}
{Cardiel} N.,  et~al., 2003, ApJ, 584, 76

\bibitem[\protect\citeauthoryear{{Christlein} \& {Zabludoff}}{{Christlein} \&
  {Zabludoff}}{2005}]{christlein2005}
{Christlein} D.,  {Zabludoff} A.~I.,  2005, ApJ, 621, 201

\bibitem[\protect\citeauthoryear{{Cram}}{{Cram}}{1998}]{cram1998}
{Cram} L.~E.,  1998, ApJL, 506, L85

\bibitem[\protect\citeauthoryear{{De Breuck}, {van Breugel}, {R{\"o}ttgering}
  \& {Miley}}{{De Breuck} et~al.}{2000}]{deBreuck2000}
{De Breuck} C.,  {van Breugel} W.,  {R{\"o}ttgering} H.~J.~A.,    {Miley} G.,
  2000, \aaps, 143, 303

\bibitem[\protect\citeauthoryear{{Doyle} et~al.,}{{Doyle}
  et~al.}{2005}]{doylemt2005mnras}
{Doyle} M.~T.,  et~al., 2005, MNRAS, 361, 34

\bibitem[\protect\citeauthoryear{{Dressler}}{{Dressler}}{1980}]{dressler1980b}
{Dressler} A.,  1980, ApJ, 236, 351

\bibitem[\protect\citeauthoryear{{G{\'o}mez} et~al.,}{{G{\'o}mez}
  et~al.}{2003}]{gomez2003}
{G{\'o}mez} P.~L.,  et~al., 2003, ApJ, 584, 210

\bibitem[\protect\citeauthoryear{{Goto}, {Yamauchi}, {Fujita}, {Okamura},
  {Sekiguchi}, {Smail}, {Bernardi} \& {Gomez}}{{Goto} et~al.}{2003}]{goto2003}
{Goto} T.,  {Yamauchi} C.,  {Fujita} Y.,  {Okamura} S.,  {Sekiguchi} M.,
  {Smail} I.,  {Bernardi} M.,    {Gomez} P.~L.,  2003, MNRAS, 346, 601

\bibitem[\protect\citeauthoryear{Hambly et~al.,}{Hambly
  et~al.}{2001a}]{hambly2001a}
Hambly N.,  et~al., 2001a, MNRAS, 326, 1279

\bibitem[\protect\citeauthoryear{Hambly et~al.,}{Hambly
  et~al.}{2001b}]{hambly2001b}
Hambly N.,  et~al., 2001b, MNRAS, 326, 1295

\bibitem[\protect\citeauthoryear{Hambly et~al.,}{Hambly
  et~al.}{2001c}]{hambly2001c}
Hambly N.,  et~al., 2001c, MNRAS, 326, 1315

\bibitem[\protect\citeauthoryear{{Hashimoto}, {Oemler}, {Lin} \&
  {Tucker}}{{Hashimoto} et~al.}{1998}]{hashimoto1998}
{Hashimoto} Y.,  {Oemler} A.~J.,  {Lin} H.,    {Tucker} D.~L.,  1998, ApJ, 499,
  589

\bibitem[\protect\citeauthoryear{{Helmboldt}, {Walterbos}, {Bothun}, {O'Neil}
  \& {de Blok}}{{Helmboldt} et~al.}{2004}]{helmboldt2004}
{Helmboldt} J.~F.,  {Walterbos} R.~A.~M.,  {Bothun} G.~D.,  {O'Neil} K.,    {de
  Blok} W.~J.~G.,  2004, \apj, 613, 914

\bibitem[\protect\citeauthoryear{{Hopkins}, {Connolly}, {Haarsma} \&
  {Cram}}{{Hopkins} et~al.}{2001}]{hopkins2001}
{Hopkins} A.~M.,  {Connolly} A.~J.,  {Haarsma} D.~B.,    {Cram} L.~E.,  2001,
  AJ, 122, 288

\bibitem[\protect\citeauthoryear{{Hopkins} et~al.,}{{Hopkins}
  et~al.}{2003}]{hopkins2003}
{Hopkins} A.~M.,  et~al., 2003, ApJ, 599, 971

\bibitem[\protect\citeauthoryear{{Hopkins}, {Schulte-Ladbeck} \&
  {Drozdovsky}}{{Hopkins} et~al.}{2002}]{hopkins2002}
{Hopkins} A.~M.,  {Schulte-Ladbeck} R.~E.,    {Drozdovsky} I.~O.,  2002, AJ,
  124, 862

\bibitem[\protect\citeauthoryear{{Hunstead}}{{Hunstead}}{1991}]{hunstead1991}
{Hunstead} R.~W.,  1991, Australian Journal of Physics, 44, 743

\bibitem[\protect\citeauthoryear{{Jones} et~al.,}{{Jones}
  et~al.}{2004}]{jonesHeath2004}
{Jones} D.~H.,  et~al., 2004, MNRAS, 355, 747

\bibitem[\protect\citeauthoryear{{Kennicutt}}{{Kennicutt}}{1998}]{kennicutt199%
8}
{Kennicutt} R.~C.,  1998, ARAA, 36, 189

\bibitem[\protect\citeauthoryear{{Lewis} et~al.,}{{Lewis}
  et~al.}{2002}]{lewis2002}
{Lewis} I.,  et~al., 2002, MNRAS, 334, 673

\bibitem[\protect\citeauthoryear{{Mart{\'{\i}}nez}, {Zandivarez},
  {Dom{\'{\i}}nguez}, {Merch{\'a}n} \& {Lambas}}{{Mart{\'{\i}}nez}
  et~al.}{2002}]{martinez2002}
{Mart{\'{\i}}nez} H.~J.,  {Zandivarez} A.,  {Dom{\'{\i}}nguez} M.,
  {Merch{\'a}n} M.~E.,    {Lambas} D.~G.,  2002, MNRAS, 333, L31

\bibitem[\protect\citeauthoryear{{Mauch},  et~al.,}{{Mauch}
  et~al.}{2003}]{mauch2003}
{Mauch} T.,     et~al., 2003, MNRAS, 342, 1117

\bibitem[\protect\citeauthoryear{{Mauch}}{{Mauch}}{2005}]{mauch2005PhD}
{Mauch} T.,  2005, PhD thesis, University of Sydney

\bibitem[\protect\citeauthoryear{Meyer et~al.,}{Meyer
  et~al.}{2004}]{meyer2004}
Meyer M.,  et~al., 2004, MNRAS, 350, 1195

\bibitem[\protect\citeauthoryear{{Mirabel} \& {Sanders}}{{Mirabel} \&
  {Sanders}}{1988}]{mirabel1988}
{Mirabel} I.~F.,  {Sanders} D.~B.,  1988, ApJ, 335, 104

\bibitem[\protect\citeauthoryear{{Nuijten}, {Simard}, {Gwyn} \&
  {R{\o}ttgering}}{{Nuijten} et~al.}{2005}]{nuijten2005}
{Nuijten} M.~J.~H.~M.,  {Simard} L.,  {Gwyn} S.,    {R{\o}ttgering} H.~J.~A.,
  2005, ApJL, 626, L77

\bibitem[\protect\citeauthoryear{{Oort}, {Steemers} \& {Windhorst}}{{Oort}
  et~al.}{1988}]{oort1988}
{Oort} M.~J.~A.,  {Steemers} W.~J.~G.,    {Windhorst} R.~A.,  1988, \aaps, 73,
  103

\bibitem[\protect\citeauthoryear{{Sodr{\'e}} \& {Mateus}}{{Sodr{\'e}} \&
  {Mateus}}{2003}]{sodre2003}
{Sodr{\'e}} L.~J.,  {Mateus} A.~J.,  2003, in ASP Conf. Ser. 297: Star
  Formation Through Time The environmental dependence of star formation in the
  nearby universe.
p.~221

\bibitem[\protect\citeauthoryear{{Stevens}}{{Stevens}}{2005}]{stevens2005}
{Stevens} J.~B.,  2005, PhD thesis, University of Melbourne

\bibitem[\protect\citeauthoryear{{Yun}, {Reddy} \& {Condon}}{{Yun}
  et~al.}{2001}]{yun2001}
{Yun} M.~S.,  {Reddy} N.~A.,    {Condon} J.~J.,  2001, ApJ, 554, 803

\bibitem[\protect\citeauthoryear{{Zwaan} et~al.,}{{Zwaan}
  et~al.}{2004}]{zwaan2004}
{Zwaan} M.,  et~al., 2004, MNRAS, 350, 1213

\bibitem[\protect\citeauthoryear{{Zwaan}, {Meyer}, {Staveley-Smith} \&
  {Webster}}{{Zwaan} et~al.}{2005}]{zwaan2005}
{Zwaan} M.~A.,  {Meyer} M.~J.,  {Staveley-Smith} L.,    {Webster} R.~L.,  2005,
  MNRAS, 359, L30

\end{thebibliography}

\section*{SUPPLEMENTARY MATERIAL}

This paper is typeset from a T$_E$X/L$^A$T$_E$X file prepared by the author.

\end{document}